\documentclass[a4paper]{article}
\pdfoutput=1 

\usepackage{jcappub} 

\usepackage[T1]{fontenc} 
\usepackage[utf8]{inputenc}

\usepackage{amsmath}
\usepackage{amssymb}
\usepackage{bm}
\usepackage{bbold}
\usepackage[compat=1.0.0]{tikz-feynman}

\usepackage{graphicx}
\usepackage{hyperref}
\hypersetup{colorlinks=true, linkcolor=teal, urlcolor=blue, citecolor=red}
\usepackage[american]{babel}
\usepackage{enumerate}
\usepackage{xcolor,colortbl}
\usepackage{mathtools}
\usepackage{subcaption}
\usepackage{soul}
\setstcolor{red}
\usepackage{makebox}
\newcommand{\U}[1]{\mathrm{U}(1)_{\mathrm{#1}}}			
\newcommand{\SU}[2]{\mathrm{SU}(#1)_{\mathrm{#2}}}		

\renewcommand{\(}{\left(}
\renewcommand{\)}{\right)}
\renewcommand{\[}{\left[}
\renewcommand{\]}{\right]}
\newcommand{\del}{\partial}
\newcommand{\abs}[1]{\left| #1 \right| }
\newcommand{\mean}[1]{\left \langle #1 \right \rangle }
\newcommand{\vev}[0]{VEV}

\usepackage{pifont}
\newcommand{\xmark}{\ding{55}}%

\newcommand\varpm{\mathbin{\vcenter{\hbox{%
  \oalign{\hfil$\scriptstyle\hspace{-0.2ex}+\hspace{-0.2ex}$\hfil\cr
          \noalign{\kern-.5ex}
          $\scriptscriptstyle({-})$\cr}%
}}}}

\newcommand\varmp{\mathbin{\vcenter{\hbox{%
  \oalign{\hfil$\scriptstyle\hspace{-0.2ex}-\hspace{-0.2ex}$\hfil\cr
          \noalign{\kern-.5ex}
          $\scriptscriptstyle({+})$\cr}%
}}}}

\usepackage[capitalise]{cleveref}

\crefname{section}{Sec.}{Secs.}
\crefname{table}{Tab.}{Tabs.}
\crefname{figure}{Fig.}{Figs.}
\crefname{equation}{Eq.}{Eqs.}
\crefname{appendix}{Appendix\ }{Appendix\ }

\graphicspath{{figures/}}

\allowdisplaybreaks

\title{Ultralight bosons for strong gravity applications from simple Standard Model extensions}

\author[a,b]{Felipe F. Freitas}
\author[c,b]{Carlos A.~R.~Herdeiro}
\author[a,b]{Ant\'onio~P.~Morais}
\author[d]{Ant\'onio Onofre}
\author[e]{Roman~Pasechnik}
\author[c,b]{Eugen Radu}
\author[c,b]{Nicolas Sanchis-Gual}
\author[f,g]{Rui Santos}

\affiliation[a]{Departamento de F\'{i}sica da Universidade de Aveiro,\\ 
	Campus de Santiago, 3810-183 Aveiro, Portugal.}
\affiliation[b]{Centre  for  Research  and  Development  in  Mathematics  and  Applications  (CIDMA),\\ 
	Campus de Santiago, 3810-183 Aveiro, Portugal.}
\affiliation[c]{Departamento de Matem\'{a}tica da Universidade de Aveiro,\\ 
	Campus de Santiago, 3810-183 Aveiro, Portugal.}
\affiliation[d]{Centro de F\'{i}sica das Universidades do Minho e do Porto (CF-UM-UP),\\
	 Universidade do Minho, 4710-057 Braga, Portugal.}
\affiliation[e]{Department of Astronomy and Theoretical Physics, Lund University,\\ 
	221 00 Lund, Sweden.}
\affiliation[f]{ISEL -  Instituto Superior de Engenharia de Lisboa,\\
	Instituto Polit\'ecnico de Lisboa  1959-007 Lisboa, Portugal.}
\affiliation[g]{Centro de F\'{i}sica Te\'{o}rica e Computacional, Faculdade de Ci\^{e}ncias,Universidade de Lisboa,\\
	 Campo Grande, Edif\'{i}cio C8 1749-016 Lisboa, Portugal.}

\emailAdd{felipefreitas@ua.pt}
\emailAdd{herdeiro@ua.pt}
\emailAdd{aapmorais@ua.pt}
\emailAdd{Antonio.Onofre@cern.ch}
\emailAdd{roman.pasechnik@thep.lu.se}
\emailAdd{eugen.radu@ua.pt}
\emailAdd{nicolas.sanchis@tecnico.ulisboa.pt}
\emailAdd{rasantos@fc.ul.pt}

\abstract{We construct families, and concrete examples, of simple extensions of the Standard Model that can yield ultralight {real or} complex vectors or scalars with potential astrophysical relevance. Specifically, the mass range for these putative fundamental bosons ($\sim 10^{-10}-10^{-20}$ eV) would lead dynamically to both new non-black  hole compact objects (bosonic stars) and new non-Kerr black holes, with masses of $\sim M_\odot$ to  $\sim 10^{10} M_\odot$, corresponding to the mass range of astrophysical black hole candidates (from stellar mass to supermassive). For each model, we study the properties of the mass spectrum and interactions after spontaneous symmetry breaking, discuss its theoretical viability and caveats, as well as some of its potential and most relevant phenomenological implications {linking them to the} physics of compact objects.}

\begin{document}
	
	\maketitle
	\flushbottom

 \section{Introduction}
The problem  of Dark Matter (DM) remains one of the greatest scientific puzzles of our time.
Amongst the proposals put forward, ultralight bosonic DM has become a hot topic within the astroparticle physics and Strong Gravity communities, in particular, being suggested as a part (or even the whole) of the DM budget of the Universe. {Such \textit{fuzzy} dark matter has been most widely considered in the mass range of around $10^{-21}$ eV (or lower), to explain the dark matter halos~\cite{Conrad:2017pms,Billard:2021uyg}; for reviews, see  $e.g.$~\cite{Li:2013nal,Suarez:2013iw,Hui:2016ltb}. But over the last decade, a considerable interest has been focused on a different (larger) mass range, mostly motivated by the interesting phenomenology resulting from the interaction of such hypothetical particles with astrophysical black holes~\cite{Arvanitaki:2010sy}.}

A key question is the fundamental high energy physics (HEP) origin of such putative ultralight bosonic particles. Often, they are justified as natural consequences of the string landscape, which has been argued to originate from a \textit{string axiverse}~\cite{Arvanitaki:2009fg}. One may ask, however, if simpler and more concrete extensions of the Standard Model (SM) of particle physics yield such particles. Such constructions would be desirable to provide a broader HEP context to the models explored by the Strong Gravity community in the context of the physics of compact objects and their astrophysical implications. This paper represents a first effort towards filling in this gap, furthermore aiming at a presentation that can provide a real bridge between the HEP, astroparticle physics and Strong Gravity communities. 
 
If such ultralight bosons exist they  can clump into self-gravitating lumps. These can be Newtonian or relativistic. In the relativistic case they form a sort of ``star", that can achieve a compactness comparable to that of black holes, being described by general relativity minimally coupled to the classical field theory describing such ultralight bosons. 

For complex  fields, the relativistic solutions are  called \textit{boson stars} (for scalars~\cite{Kaup:1968zz,Ruffini:1969qy}, see reviews~\cite{Schunck:2003kk,Liebling:2012fv})  or \textit{Proca stars} (for vectors~\cite{Brito:2015pxa}). Collectively they are called \textit{bosonic stars}, regardless of their scalar/vector nature. For some comparative studies see~\cite{Herdeiro:2017fhv,Herdeiro:2020jzx} for non-rotating and~\cite{Herdeiro:2019mbz} for rotating bosonic stars.

Bosonic stars are time independent ($i.e.$ in equilibrium) spacetime geometries. In part of their parameter space, often called \textit{domain of existence}, they are perturbatively stable and can form dynamically~\cite{Seidel:1993zk} - see~\cite{Liebling:2012fv} for a review of the scalar case and~\cite{Sanchis-Gual:2017bhw,DiGiovanni:2018bvo} for the  Proca case. The analysis of stability for spinning (rather than static) bosonic stars is more  subtle. Recent studies have addressed the stability and formation of spinning bosonic stars providing evidence that there are stable solutions both for scalar and  Proca stars, with the former seeming to require appropriate self-interactions~\cite{Sanchis-Gual:2019ljs,DiGiovanni:2020ror,siemonsen2021stability}. Thus, bosonic stars are  dynamically interesting   objects, whose reality depends essentially on the existence of their fundamental  bosonic particle constituents. 

For real  fields, on the other hand, the self-gravitating  lumps are  called \textit{oscillatons}~\cite{Seidel:1993zk}. They are slightly time-dependent and decay, but they can be very long lived,  at least for spherical stars~\cite{Page:2003rd}. Thus, they can still be dynamically interesting, albeit such models have been less explored in the literature. 

The reason why one needs ultralight fields for these stars, both in the real or complex cases, is related to the requirement of obtaining stars with astrophysical masses. This will be explained quantitatively in the next section.

\medskip 
 
Gravitational wave detection offers an opportunity to  test  the  existence of such bosonic stars, within a certain mass range, for both the stars and the fundamental bosons. Over the years, waveforms for binaries of scalar and vector bosonic stars have been obtained - see e.g.~\cite{Palenzuela:2007dm,Palenzuela:2017kcg,Bezares:2017mzk,Sanchis-Gual:2018oui}, that could now be confronted with real events. Recently, moreover, the event GW190521~\cite{Abbott:2020tfl} was shown to be compatible with a collision of spinning Proca stars~\cite{CalderonBustillo:2020srq}, with a slight statistical preference with respect to the vanilla binary black  hole scenario used by the LIGO-Virgo collaboration. The fundamental boson in this scenario is a complex vector particle with mass $\mu\sim 9\times 10^{-13}$ eV. This further suggests investigating HEP scenarios where such a particle can emerge.

\medskip 

Ultralight bosons can also lead to a different sort of strong gravity signatures. They can interact with spinning black holes in the astrophysical mass range through the phenomenon of superradiance~\cite{Brito:2015oca}. This process transfers part of the black  hole rotational energy into the creation of a bosonic cloud around the horizon~\cite{Arvanitaki:2010sy}. In the case of complex scalars, this  can lead to a new type of black  holes, with scalar~\cite{Herdeiro:2014goa} or Proca hair~\cite{Herdeiro:2016tmi}. These black holes can also leave observational signatures, both in gravitational waves and other observables, like the black hole shadow~\cite{Cunha:2019ikd}.
 
 \medskip

Although challenging, searches for very low mass bosons have been performed since several decades already, to push the mass and couplings boundaries of these particles, even at the LHC. One particular example is the axion (spin=0) or axion like particle (ALP), a potential dark matter candidate that was initially proposed to explain the strong CP problem~\cite{Peccei:1977hh} and may impact the cosmology and formation of the early Universe. Given the wide variety of ALP's, many ideas have been put forward, over the past decade or so~\cite{Graham:2015ouw, Irastorza:2018dyq, Sikivie:2020zpn, Choi:2020rgn}, to probe their mass and couplings in regions not accessible before. One of the most natural ways of searching for these particles at present and future colliders, is through the couplings to Standard Model (SM) bosons~\cite{Bauer:2018uxu}. Depending on the ALP mass, the strategy changes considerably and, in the case of low mass ALP's (lower than the electron mass), couplings to photons can play an important role and may be probed at colliders, in particular at the LHC~\cite{Florez:2021zoo,Baldenegro:2018hng,Aad:2020cje,Sirunyan:2018fhl,Citron:2018lsq} through associated production or decay. Although the direct detection of photons pose severe constraints on the mass reach at colliders (few~MeV)~\cite{Baldenegro:2018hng}, indirect production of ALP is to be considered, as well.

Light vector bosons are also being searched for in precision physics at colliders. Although generically with weaker mass limits, they are worth exploring including the ones motivated by gravitational physics, once they come from laboratory experiments
and are independent of any astrophysical assumption. If these new bosons couple, even feebly, with the SM electroweak bosons through portals or hidden sectors they can, in principle, be accessible experimentally and provide crucial information in case of discovery, eventually about the nature of dark matter itself. Massless and massive (few~MeV) dark photons are examples of these type of particles that have been searched for 
(see~\cite{Fabbrichesi:2020wbt, Filippi:2020kii, Graham:2021ggy} for a
review of dark photon phenomenology and experimental tests). For massless dark photons, several precision tests set stringent limits on the dark dipole scale ~\cite{Fabbrichesi:2020wbt} $\Lambda^2/(\sqrt{\alpha_\mathrm{D}}d_\mathrm{M})$, where $\Lambda$, $\alpha_\mathrm{D}$ and $d_\mathrm{M}$ are the effective scale for the interaction, the dark photon coupling and dark photon magnetic dipole, respectively. Rare decays of $K^+$ ($K^+\rightarrow \pi^+ \nu \bar{\nu}$)~\cite{CortinaGil:2020vlo,Fabbrichesi:2020wbt} have set a lower bound (at 90\% CL) on the ratio above $\sim 9.5\times 10^6$~TeV$^2$. Although searches for massless dark photons have been neglected with respect to massive ones, the potential to search for signals in flavor physics, or in the detection of single photon events from the decays of Higgs and $Z$ bosons acompained by a dark photon, are signatures accessible at the LHC.

To conclude, we emphasise that if the ultralight states are sufficiently decoupled in both the mass and interaction spectrum, collider searches in the foreseeable future may be hopeless for detecting such states. In this case,  the (astro)physics of compact objects (and their gravitational waves), primordial gravitational  waves together with other cosmological/astrophysical phenomenology of ultralight DM, will be the only possible ways to probe the proposed models.



The paper is organized as follows. In section~\ref{Sec:SGM} we introduce the strong gravity models and in section~\ref{Sec:HEP} we introduce the high energy physics (HEP) models.  In 
section~\ref{Sec:concHEP} we discuss concrete HEP models that match the need for ultralight bosons. We summarise our findings in section~\ref{sec:Conc}.

 \section{Strong Gravity models}
 \label{Sec:SGM}
 Let us start with a brief description of the classical field theory models that are commonly in use by the gravity community to address the gravitational effects of such putative ultralight bosons.
 
Einstein's gravity in 3+1 dimensional spacetime is
minimally coupled to a spin-$s$ field, where $s$ takes one of the two values: $s=0$ or $s=1$.
The action is (with $c=1$)
\begin{eqnarray}
\label{action}
\mathcal{S}=\int d^4 x \sqrt{-g}
\left [
\frac{R}{16 \pi G}
+
\mathcal{L}_{(s)}
\right] \ ,
\end{eqnarray}
where $R$  is the Ricci scalar of  the   spacetime metric $g$. The $s=0$ (scalar) and $s=1$ (Proca) matter Lagrangians are:
\begin{eqnarray}
\label{LS}
&& \mathcal{L}_{(0)}= - g^{\alpha \beta}\bar \Phi_{, \, \alpha} \Phi_{, \, \beta} - \mu^2 \bar \Phi \Phi-V_{(0)}^{\rm  int}(\bar \Phi \Phi) \ , \\
&& \mathcal{L}_{(1)}= -\frac{1}{4}\mathcal{F}_{\alpha\beta}\bar{\mathcal{F}}^{\alpha\beta}
-\frac{\mu^2}{2}\mathcal{A}_\alpha\bar{\mathcal{A}}^\alpha-V_{(1)}^{\rm  int}(\mathcal{A}_\alpha\bar{\mathcal{A}}^\alpha) \ .
\label{LD}
\end{eqnarray}
Here, $\Phi$ is a complex scalar field;
$\mathcal{A}$ is a complex 4-potential, with the field strength $\mathcal{F}_{\alpha\beta} =\partial_\alpha\mathcal{A}_\beta-\partial_\beta\mathcal{A}_\alpha$.
In both cases, $\mu>0$ corresponds to the mass of the field(s).
The overbar denotes complex conjugation and $V_{(s)}^{\rm  int}$ describes the self-interactions term in each  case. One can, of course, consider also generalised models including possibly many scalar and/or vector states with different masses, as well as non-minimal couplings between them and/or with gravity. Additionally (or alternatively) one can consider real fields, with Lagrangians identical to eqs.~\eqref{LS} and~\eqref{LD} but dropping the overbars. Thus, one can face the model~\eqref{action} as the simplest (and in a way the basic buliding block), but not the most general field theory that could have interesting phenomenology for compact  objects.

\subsection{Maximal mass of bosonic stars}
\label{secmaxmass}
The simple models given by eqs.~\eqref{LS} and~\eqref{LD} yield a valuable lesson which justifies the focus on ultralight particles. 
The maximum  Arnowitt-Deser-Misner (ADM)  mass of a scalar boson star made up of a free complex scalar field ($V_{(0)}^{\rm  int}=0$) is of the order of the Compton wavelength of the scalar field:
\begin{equation}
 M_{\rm ADM}^{\rm max} = \alpha_{\rm BS}^{(s)} \frac{M_{\rm Pl}^2}{\mu} = \alpha_{\rm BS}^{(s)} \, 1.34\times 10^{-19}M_\odot\left(\frac{\rm GeV}{\mu}\right)\ ,
\label{mini}
\end{equation}
where $M_{\rm Pl},M_{\odot}$ denote the Planck mass and solar mass, respectively. The constant $\alpha_{\rm BS}^{(s)}$ is obtained from computing the explicit solutions and it is of order unity, but its specific value depends on the quantum numbers of the boson star. For instance, for the fundamental and most stable, spherically symmetric stars $\alpha_{\rm BS}^{(0)}=0.633$~\cite{Liebling:2012fv}, whereas for the fundamental rotating stars  $\alpha_{\rm BS}^{(0)}=1.315$~\cite{Yoshida:1997qf,Grandclement:2014msa}. Thus, for typical SM particle masses, say $\mu\sim 1$ GeV, this maximal mass is small, by astrophysical standards (and particularly so concerning the known compact objects):\footnote{There are, however, speculative compact objects such as primordial black holes that could have sub-solar masses.} $M_{\rm ADM}^{\rm max}\sim 10^{-19} M_\odot$. For this reason such boson stars were historically dubbed \textit{mini-}boson stars. On   the other  hand, if one allows for ultralight particles, in the  mass range
 \begin{equation}
     \mu \ \in \ [10^{-20},10^{-10}] \ \mathrm{eV} \ ,
     \label{massin}
 \end{equation}
 the maximal mass is in the range of the known  astrophysical black holes
  \begin{equation}
  M_{\rm ADM}^{\rm max}  \  \in \ [1,10^{10}] \ M_{\odot} \ .
  \label{massbhs}
 \end{equation}
 
Self-interactions ($V_{(0)}^{\rm  int}\neq 0$) can change the relation between the fundamental boson mass $\mu$ and the maximal mass a boson star  can support $M_{\rm ADM}^{\rm max}$~$e.g$~\cite{Colpi:1986ye,Herdeiro:2015tia}. {We will further comment on this in \cref{sec:real-scalar}, but for the moment we will take the simplest free models described above as the motivation to consider ultralight bosons.} 
 
For Proca stars without self-interactions ($V_{(1)}^{\rm  int}=0$), the relation~\eqref{mini} also holds and  only the values of the constant $\alpha_{\rm BS}^{(s)}$ differ. For instance, for the fundamental and most stable, spherically symmetric stars $\alpha_{\rm BS}^{(1)}=1.058$~\cite{Brito:2015pxa,Herdeiro:2017fhv}, whereas for the fundamental rotating stars  $\alpha_{\rm BS}^{(1)}=1.125$~\cite{Herdeiro:2019mbz}.
 
The bottom line is that the scaling of the maximal mass with the inverse of the boson mass, $cf.$ \cref{mini}, requires ultralight particles in order for bosonic stars to have the masses of astrophysical black holes. Even lighter bosonic particles are required  for the description  of galactic DM halos~\cite{Li:2013nal,Suarez:2013iw,Hui:2016ltb}, since  they correspond to larger {length scales (via their Compton wavelength) which sets the length scale of the gravitational structures they form.} 

\subsection{Superradiance and ``hairy" black holes}

A second reason (partly related to the first) why the mass interval \eqref{massin} is interesting for strong gravity systems, relates to the interaction of such putative ultralight bosons with black holes. 

The rotational energy of \textit{spinning} black holes can be mined by a classical process called superradiance - see~\cite{Brito:2015oca} for a comprehensive review.\footnote{The rotational energy of black holes is thought to power significant astrophysical events, such as powerful jets in Active Galactic Nuclei (AGNs) and other active galactic centres.} This process can be mediated by ultralight bosonic particles. That is, bosonic  modes with frequency $\omega\in \mathbb{R}^+$ and azimuthal harmonic index $m\in \mathbb{N}$ (which could  be, e.g., quantum fluctuations) will be amplified if they have frequency in the  superradiant regime, $0<\omega<m\Omega_H$, where $\Omega_H$ is the horizon angular  velocity of a Kerr black hole (see e.g.~\cite{Townsend:1997ku}). These modes grow, exciting many bosonic  quanta in the same state, becoming a Bose-Einstein condensate around the spinning black hole, which spins down as a result of the transfer of energy and angular momentum to the bosonic cloud around  it.  The process stalls once the  black hole horizon angular velocity slows down enough to meet the  phase angular velocity of the dominant growing mode, $\omega/m$,  as fully non-linear numerical simulations have shown~\cite{Sanchis-Gual:2015lje,Bosch:2016vcp,East:2017ovw}.  The new equilibrium point is a black hole with bosonic hair~\cite{Herdeiro:2017phl} (scalar~\cite{Herdeiro:2014goa} or Proca~\cite{Herdeiro:2016tmi,Santos:2020pmh}),  when the bosonic field is complex. These black holes can have an interesting and distinct  phenomenology. If the bosonic field is real, on the other hand, the new equilibrium state  is non-stationary, and the  bosonic cloud decays by emitting gravitational waves, since  the bosonic cloud's energy distribution  is non-axisymmetric~\cite{Arvanitaki:2010sy}. Current LIGO-Virgo-KAGRA searches aim at setting bounds on such putative continuum backgrounds of gravitational waves~\cite{Zhu:2020tht}.

In this scenario, the importance of the boson mass is related to the \textit{efficiency} of the superradiance process. Although bosonic modes with any mass may trigger the energy extraction process,  the timescale for the runaway process that grows a macroscopic cloud depends sensitively on a resonance betweeen the Compton wavelength of the bosonic particle and the Schwarzschild radius of the black hole. Away from this sweet spot the time scale grows very fast~\cite{Brito:2015oca}. Thus, efficient  superradiant energy extraction from astrophysical black holes, which have the mass range~\eqref{massbhs}, requires bosonic particles in the range given in~\eqref{massin}.

 \section{HEP models - general principles for ultralight bosons}
 \label{Sec:HEP}
 
Having motivated that hypothetical ultralight bosonic particles are interesting from the viewpoint of strong gravity systems, we turn to the question of their HEP origin. What goes into models that may have such incredibly small masses? Fine-tuning? Then, how would such masses be protected against quantum corrections? Additionally, for such particles to evade current collider constraints, couplings to the SM particles must be zero or  very small. How is this consistently accomplished? These are some of the obvious questions that must be addressed. In this section we shall discuss different generic principles. Concrete models will follow in the subsequent sections.
 
The emergence of ultralight bosonic particles in HEP models can be obtained in a number of different ways. These can either be scalars (spin-0) or vectors (spin-1), and the nature of their origin is often intrinsically related to the type of symmetries present in a given model.

Let us first simply assume \textit{fine-tuning}: that a tiny physical mass of a given scalar emerges due to a remarkable cancellation of theory parameters that Nature might have accidentally picked. Independently of their size such a cancellation can be fine-tuned to the desired degree. However, this rather crude solution lacks from a fundamental explanation and is typically plagued with severe problems. In particular, scalar masses are unprotected against quantum corrections and can receive large contributions. For example, if the fine-tuned ultralight mass, say of the order $\mathcal{O}(10^{-10}~\mathrm{eV})$, is well below the natural energy-scale of the theory, for instance the Higgs boson mass or the electroweak (EW) scale, $\mu_\mathrm{EW} \sim \mathcal{O}(100~\mathrm{GeV})$, the dominant quantum corrections are typically of the order $\mathcal{O}(1~\mathrm{GeV})$ completely spoiling the original mass. Quantum corrections to a particle mass can be expanded in a perturbative series. Therefore, to preserve the size of the uncorrected mass, it is necessary to rely on an order-by-order cancellation which, although possible, seems unlikely.

As an alternative to the fine-tuned solution one might consider a \textit{feebly coupled} theory. In other words, the size of an ultralight scalar does not result from a cancellation of large parameters, but instead, it is proportional to a set of tiny theory parameters  that compensate for any sizeable effects from the much larger EW scale. Furthermore, quantum corrections can be kept under control provided that the same parameters that suppress the mass will also suppress all quantum effects. A feebly interacting sector is effectively decoupled from the EW scale. Therefore, besides a gravitational footprint, it may become rather challenging to search for complementary signatures of such a theory in collider experiments. 

At this point, the question that one might pose is if there is any way of simultaneously allowing for an ultralight scalar without imposing tiny couplings. The answer is yes and can be achieved based on \textit{symmetry} principles.

Let us then consider an alternative scalar sector equipped with a certain global continuous symmetry. According to the \textit{Goldstone's Theorem} \cite{Peskin:1995ev}, if such a symmetry is spontaneously broken there will be as many massless scalar degrees of freedom as the number of broken symmetry generators. These are typically dubbed as Goldstone bosons and, for the case of a global symmetry, they remain in the physical particle spectrum. At this point, it is evident that, independently of the energy scale at which this breaking occurs, such a model can naturally offer massless scalars. However, these are neither candidates for building up bosonic stars (as they are massless), nor their cosmology is favoured due to stringent constraints on the production of excessive dark radiation \cite{Planck:2015fie}. In HEP, the typical approach is to softly break such a symmetry. By soft breaking we mean that the size of such explicit breaking is small in comparison to the scale of the theory and, most importantly, quantum effects do still preserve the original symmetry to all orders in perturbation theory. As a result, the Goldstone bosons acquire a finite mass that can be arbitrarily small and is protected against quantum corrections. Notice that the interaction strength of these pseudo-Goldstone particles with the Higgs boson, or even other heavy scalars that the model might predict, can be sizeable enough to be at the reach of collider experiments.

So far we have only discussed possibilities for spin-0 bosons. However, HEP models can also consistently predict ultralight Proca fields. In particle physics, the gauge principle forbids explicit mass terms for vector bosons which can only be generated upon spontaneous breaking of a continuous gauge symmetry. We can then regard this class of models as an extension of the latter by promoting the global symmetry to a local one. The question now is how can one generate such an incredibly small mass to a Proca field? The answer is twofold. First, we can rely, once again, in a feebly interacting gauge sector. In other words, given that the Proca field mass, $\mu_\mathrm{Proca} \propto g v$, is proportional to a new gauge coupling $g$ and the scale at which the breaking occurs $v$, one can always choose $g$ to be as small as we might need such that $g v \lesssim \mathcal{O}(10^{-10}~\mathrm{eV})$. However, in this limit we are decoupling the Proca field to a level where it becomes essentially non-interacting, thus invisible, except by its gravitational effect. In fact, no matter how small the couplings are, the equivalence principle tells us that all matter/energy gravitates and thus one cannot make any particle invisible to gravity. Alternatively, one can rely on a tiny breaking scale such that $v/\mu_\mathrm{EW}$ is very small. As such, the Proca field can become ultralight independently of the value of $g$. This is a rather elegant hypothesis as the size of the Proca field can be seen as a signature of a, yet to discover, new-physics (NP) scale, well below the EW one. Last but not least, it is well known that vector boson masses are protected against quantum corrections by virtue of the gauge symmetry \cite{Peskin:1995ev}.

\section{Concrete HEP models}
 \label{Sec:concHEP}
We now turn to concrete models where ultralight bosons may emerge. Our aim is to present realistic benchmark scenarios; thus the Higgs and the EW gauge bosons must be included. It is then instructive to start by revisiting the bosonic EW sector of the SM. First, let us introduce the scalar (Higgs) potential
\begin{equation}
\label{eq:VSM}
V_0\(H\)= \mu^2_H H^\dagger H + \frac12\lambda_H (H^\dagger H)^2\,,
\end{equation}
which is manifestly invariant under $\SU{2}{W} \times \U{Y}$ transformations, dubbed the EW symmetry, and where W refers to weak interactions whereas Y denotes the weak hypercharge. In \eqref{eq:VSM} $H$ is the SM Higgs $\SU{2}{W}$ doublet whose real valued components can be written as
\begin{equation}
\begin{aligned}
H = \dfrac{1}{\sqrt{2}} 
\begin{pmatrix}
\omega_1 + i \omega_2  \\
v_h + h + i z
\end{pmatrix}\,.	
\end{aligned}
\label{eq:H}
\end{equation}
While $v_h$ is the vacuum expectation value (\vev) that describes the classical ground state configurations of the theory, $h$ represents radial quantum fluctuations around such a minimum of the potential \eqref{eq:VSM}. The Goldstone modes $\omega_{1,2}$ and $z$ are absorbed by longitudinal degrees of freedom of the $W^\pm$ and $Z$ gauge bosons once the EW symmetry is spontaneously broken by the Higgs doublet \vev~$\mean{H} = \tfrac{1}{\sqrt{2}}\begin{pmatrix}
0 && v_h
\end{pmatrix}^\top$,
with $v_h \approx 246~\mathrm{GeV}$. The coupling of the EW vector bosons with the Higgs doublet is described by the following kinetic terms
\begin{equation}
	\mathcal{L}_\mathrm{kin}^0 \supset D_\mu H^\dagger D^\mu H~\text{with}~ D_\mu = \partial_\mu + i g_1 Y B_\mu + i g_2 \frac{\tau_a}{2} A_\mu^a \ ,
	\label{eq:Lkin}
\end{equation}
where $D_\mu$ is the covariant derivative, $g_{1,2}$ are the $\U{Y}$ and $\SU{2}{W}$ gauge couplings, $\tau_a$ ($a = 1,2,3$) the Pauli matrices and $B_\mu$, $A_\mu^a$ the massless electroweak gauge bosons. In what follows, such a bosonic EW-sector of the SM should be implicit.
 
Before proceeding, however,  we would like to mention the existence of a solitonic solution of \textit{just} the above (bosonic) EW equations of motion taken from $\mathcal{L}_\mathrm{kin}^0$ and $V_0\(H\)$ (thus, in the spirit of the aforementioned bosonic stars, but without the need of hypothetical new particles)
-- the \textit{sphaleron} \cite{Manton:1983nd}.  This non-perturbative classical solution relies on a balance between the scalar (Higgs) and (non-Abelian) gauge interactions. In particular, it has been suggested as a mechanism for a possible explanation of baryonic asymmetry~\cite{Manton:1983nd,Klinkhamer:1984di}.
The sphaleron is usually studied neglecting gravity ($i.e.$ in Minkowski spacetime); its self-gravity does not change significantly its properties
\cite{Volkov:1998cc}. But considering gravity, one can build a non-linear superposition between a sphaleron and a black hole horizon, constructing a black hole with sphaleron hair.\footnote{The same occurs for spinning bosonic stars: one can place a black hole horizon at their centre, leading to the hairy black holes in~\cite{Herdeiro:2014goa,Herdeiro:2016tmi,Santos:2020pmh}.} This is an interesting counter example of the black hole no hair conjecture \cite{Greene:1992fw}. These configurations,  however, are unstable
\cite{Mavromatos:1995kc,Winstanley:1995iq}; furthermore, they are likely limited to the microscopic realm, with an ADM mass $\sim$ 10 TeV, thus unlikely to be relevant in an astrophysical context. But we note that the mechanism allowing for the sphaleron's existence may also work in the various SM extensions considered below, with a crucial role played by the non-Abelian interactions.

\subsection{Ultralight scalars: a minimal approach}
\label{sec:uls}

\subsubsection{Fine-tuned $vs.$ a feebly coupled theory: the case of complex scalars}
\label{sec:FT}

\begin{figure}[]
    \centering
    \includegraphics[width=0.3\textwidth]{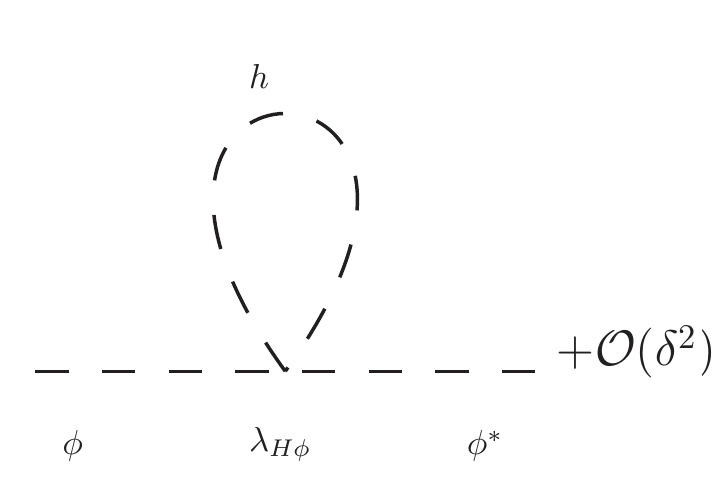}
	\caption{Leading, one-loop, quantum corrections to the $m_\phi^2 \phi^\ast \phi$ quadratic term in the mass basis: $\delta = \tfrac{1}{16 \pi^2}$ is the perturbative expansion parameter.}
	\label{fig:FT}
\end{figure}

The first model to consider is among the simplest extensions of the SM that one might think of. It consists of an additional complex singlet $\phi$ charged under a global $\U{G}$ symmetry such that the theory is invariant under the phase transformation
\begin{equation}
    \phi \to e^{i \alpha} \phi\, .
\end{equation}
The scalar potential before electroweak symmetry breaking (EWSB) reads as
\begin{equation}
	V\(H,\phi\) = V_0\(H\) + \mu_\phi^2 \phi^\ast \phi + \frac{1}{2} \lambda_\phi \abs{\phi^\ast \phi}^2
	+ \lambda_{H \phi} H^\dagger H \phi^\ast \phi \ ,
	\label{eq:V1}
\end{equation}
where $\lambda_{H \phi}$ is typically dubbed as the \textit{Higgs portal coupling} since it is the only interaction where the new scalar can couple to the SM. Note that the potential $V\(H,\phi\)$ is bounded from below whenever the conditions
\begin{equation}
	\lambda_H,\lambda_\phi > 0 \qquad, \qquad \lambda_H \lambda_\phi - \lambda_{H \phi}^2 > 0\,,
\end{equation}
are verified. Expanding the theory around the minimum of the potential
one obtains the condition $\mu_H^2 = -\lambda_H v_h^2$. Replacing $\mu_H^2$ in \eqref{eq:V1}, the Hessian matrix evaluated in the vacuum of the theory comes already in a diagonal form and reads as
\begin{equation}
\bm{M}^2 =
\begin{pmatrix}
\bm{0}_{3 \times 3} & \bm{0}_{3 \times 1} & \bm{0}_{3 \times 1} & \bm{0}_{3 \times 1} \\ 
\bm{0}_{1 \times 3} & 2 \lambda_H v_h^2 & 0 & 0 \\
\bm{0}_{1 \times 3} & 0 & \mu_\phi^2 +\tfrac12 \lambda_{H \phi}
v_h^2 & 0 \\
\bm{0}_{1 \times 3} & 0 & 0 & \mu_\phi^2 +\tfrac12 \lambda_{H \phi} v_h^2 
\end{pmatrix}\,,
\label{eq:hess1}
\end{equation} 
where we identify the SM Higgs boson and new complex scalar masses as
\begin{equation}
	m_h^2 = 2 \lambda_H v_h^2 \qquad m_\phi^2 = \mu_\phi^2 +\tfrac12 \lambda_{H \phi} v_h^2\,,
	\label{eq:spec1}
\end{equation}
respectively. The relevant cubic and quartic self interactions involving the singlet $\phi$ are trivially given as
\begin{equation}
    \lambda_{_{h \phi \phi}} = v_h \lambda_{H\phi}\,, \qquad \lambda_{_{h h \phi \phi}} = \lambda_{H \phi}\,, \qquad \lambda_{_{\phi \phi \phi \phi}} = \lambda_{\phi}\,.
\end{equation}
While the mass of the Higgs boson is well known to be $m_h \approx 125~\mathrm{GeV}$, the model has enough freedom to allow for an ultralight complex scalar in one of the following cases:
\begin{enumerate}
	\item \textbf{Fine-tuned scenario:} For a portal coupling of order $\mathcal{O}(1)$, if we require $\mathrm{Sign}(\mu_\phi^2) = - \mathrm{Sign}(\lambda_{H \phi})$ and $\abs{\mu_\phi^2} \approx \tfrac12 \abs{\lambda_{H \phi}} v_h^2$ such that they differ by no more than one part in $10^{20}$, then $m_\phi \lesssim 10^{-10}~\mathrm{eV}$ can be achieved.
	\item \textbf{Feebly interacting scenario:} Alternatively, if we now allow the portal coupling to be tiny, $i.e.$ $ 10^{-62} \lesssim \lambda_{H \phi} \lesssim 10^{-42}$ and now take $\mu_\phi^2$ to be of the same order of $\lambda_{H \phi} v_h$, then one can also have $10^{-20} \lesssim m_\phi \lesssim 10^{-10}~\mathrm{eV}$. 
\end{enumerate}
The first scenario relies on a remarkable fine-tuning of quantities that are of the size of the EW scale. While at first glance it may seem an easy choice, the inherent complications that come together with such a solution are rather unattractive. In particular, an ultralight scalar at least 21 orders of magnitude below the EW scale poses a tremendous hierarchy problem which is rather more severe than the well known Higgs boson mass hierarchy problem in the SM \cite{Susskind:1978ms,Gildener:1976ai}. To give an idea of what is involved, let us consider that $\lambda_{H \phi} \gtrsim \lambda_\phi \sim \lambda_H$ such that the dominant quantum corrections to the mass $m_\phi^2$ are coming from portal interactions, given at leading order by the Feynman diagram in \cref{fig:FT}. 
These can be translated into the correction
\begin{equation}
    M_\phi^2 = m_\phi^2 + \Delta^2 \qquad \textrm{with} \qquad \Delta^2 \sim \mathcal{O}\left( \delta \lambda_{H \phi} \mu_\mathrm{EW}^2 \right)\,,
    \label{eq:corr}
\end{equation}
and where the loop factor is $\delta = 1/16 \pi^2$. Recalling that the EW scale is of the order $10^{11}~\mathrm{eV}$, it becomes clear from \cref{eq:corr} that the first order term correction is such that, at least, a cancellation degree of $\Delta^2/m_\phi^2 \approx \delta \lambda_{H \phi} 10^{31} \approx 10^{29}$ is necessary. As one goes to higher order terms the degree of cancellation gets relaxed by increasing powers of $\delta \lambda_{H \phi}$. However, the size of $m_\phi^2$ can only be preserved if all such loop orders cancel simultaneously.

It is clear from the discussion above that the minimal way to control the size of $\Delta$ is in the limit of a feebly interacting theory where $\lambda_{H \phi} \to 0^+$. The portal coupling must be extremely small, still non-zero, in order to compensate for the size of the EW scale and simultaneously suppress $\Delta$ to a level where quantum corrections are no longer an issue. The mass parameter $\mu_\phi^2$ also needs to be of the same order as the product $\lambda_{H \phi} v_h^2$. Notice that in this scenario the presence of such an ultralight scalar may only be probed in a gravitational channel provided that the coupling to the Higgs boson, and any other SM particle, is effectively zero.

\subsubsection{Generalization to larger multiplicities}
\label{sec:mult}

The potential in \cref{eq:V1} can be easily generalized to the case of a complex scalar that transforms as a fundamental representation of a global $\SU{N}{}$ group. In particular, one could define a new $\bm{\Phi}$ field
\begin{equation}
\begin{aligned}
\bm{\Phi} = \dfrac{1}{\sqrt{2}} 
\begin{pmatrix}
\rho_1 + i \pi_1  \\
\vdots \\
\rho_N + i \pi_N
\end{pmatrix}\, , 	
\end{aligned}
\label{eq:SUNphi}
\end{equation}
such that the new scalar potential is identical to that in \cref{eq:V1} but with the replacements $\phi \to \bm{\Phi}$ and $\phi^\ast \to \bm{\Phi}^\dagger$. The same discussion regarding a fine-tuned or a feebly interacting theory applies, the mass spectrum and self couplings are also identical to the ones above and the only distinction resides in the multiplicity of equal mass complex fields, in particular
\begin{equation}
    m_{\phi_1}^2 = m_{\phi_2}^2 = \cdots = m_{\phi_N}^2 \qquad \textrm{with} \qquad \phi_i = \frac{1}{\sqrt{2}} (\rho_i + i \pi_i)\,.
    \label{eq:SUNmass}
\end{equation}

The $\SU{N}{}$ symmetry forces an equal mass to every complex scalar as seen in \cref{eq:SUNmass}, which may be too restrictive. However, in a completely generic setting without any symmetry, the amount of possible interactions between $N$ different scalars is enormous, substantially complicating the problem. A simpler solution leading to $N$ non-degenerate masses is to replace the $\SU{N}{}$ by a smaller symmetry as $e.g.$~$\U{1} \times \cdots \times \U{N}$ where each of the $N$ scalars is charged under a different $\U{}$. The advantage of this structure is that the potential can be constructed solely in terms of the operators $\phi^\ast_i \phi^i$ such that a generalization of \cref{eq:V1} reads as
\begin{equation}
	V\(H,\phi\) = V_0\(H\) + \mu_i^2 \phi_i^\ast \phi^i + \frac{1}{2} \lambda_{ij} \phi_i^\ast \phi^i \phi_j^\ast \phi^j
	+ \lambda_{H i} H^\dagger H \phi^\ast_i \phi^i\,.
	\label{eq:Vmulti}
\end{equation}
The $N$ masses are now non-degenerate and equal to
\begin{equation}
    m_{\phi_i}^2 = \mu_i^2 + \frac{1}{2} \lambda_{Hi} v_h^2\,.
\end{equation}

As a side note, let us remark that, from the viewpoint of gravity, some of the effect of considering $N$ complex scalars can be understood by simple arguments. For instance, consider a (mini-)boson star composed by such $N$ complex scalars with the same mass and without self-interactions. How does its maximal mass~\eqref{mini} change? The corresponding action
\begin{equation}
	\label{action2}
	\begin{aligned}
	\resizebox{0.9\columnwidth}{!}{%
$\mathcal{S}=\int d^4 x \sqrt{-g}
\left [
\frac{R}{16 \pi G}- g^{\alpha \beta} \phi_{i, \, \alpha}^\ast \phi_{, \, \beta}^i - \mu_i^2 \phi_i^\ast \phi^i
\right]
\stackrel{\phi_i\equiv\phi}= N \int d^4 x \sqrt{-g}
\left [
\frac{R}{16 \pi G_{\rm efe}}- g^{\alpha \beta} \phi_{, \, \alpha}^\ast \phi_{, \, \beta} - \mu^2 \phi^\ast \phi
\right] \ ,$
}
	\end{aligned}
\end{equation}
becomes that of a single complex field but with an effective Newton's constant $G_{\rm efe}\equiv GN$ under the assumption that all fields acquire the same classical profile $\phi_i\equiv\phi$. This implies the maximal mass scales \textit{down} with N:
\begin{equation}
 M_{\rm ADM}^{\rm max} = \alpha_{\rm BS}^{(s)} \frac{M_{\rm Pl}^2}{N\mu} = \alpha_{\rm BS}^{(s)} \, 1.34\times 10^{-19}M_\odot\left(\frac{\rm GeV}{N\mu}\right)\ ,
\label{Nmini}
\end{equation}
where $\alpha_{\rm BS}^{(s)}$ are the same numbers discussed in Section~\eqref{secmaxmass}. Thus, allowing a multiplicity of the fields does not help in getting astrophysical masses with heavier bosons.

\subsubsection{Ultralight real scalars}

So far we have only discussed the possibility of complex scalars which are relevant to describe astrophysical objects such as boson stars or black holes in equilibrium with scalar clouds. However, real scalars are also known to offer an interesting phenomenology in the form of continuous monochromatic gravitational waves that dissipate the angular momentum of the black hole-scalar cloud  system~\cite{east2018massive,siemonsen2020gravitational} and could be detected by current searches to (dis)prove the existence of bosonic clouds around spinning black holes~\cite{brito2017gravitational,palomba2019direct}, if such bosonic particles exist in Nature. Besides, real fields can also form massive solitonic solutions known as oscillatons that may be very long lived and have potential astrophysical interest~\cite{seidel1991oscillating,Page:2003rd,brito2016interaction,helfer2019gravitational,widdicombe2020black}. 

If we consider the case of a single real scalar $\varphi$ feebly interacting with the Higgs boson, the discussion is in many aspects identical to what we have already argued above. The main difference is the lack of a Noether charge that allows new interactions of the form $\varphi H^\dagger H$ and $\varphi^3$. However, while the latter plays no role on the mass, the former is a portal-like coupling and needs to be extremely small in order to prevent large contributions. One possible solution to prevent such an effect is to impose a discrete $\mathbb{Z}_2$ symmetry where $\varphi \to - \varphi$ which would forbid both terms and result in a mass spectrum analogous to that of the complex case.

\subsection{Softly broken global symmetry hypothesis}
\label{sec:soft}

While the fine-tuning solution introduces a severe hierarchy problem, the choice of extremely small portal couplings is, to some extent, arbitrary. Furthermore, the effect of an incredibly small interaction can be ignored and simply consider that the $\mu_\phi^2$ parameter describes, on its own, the mass of the new boson. However, one can ask whether tiny masses, protected against quantum corrections, can emerge as a result of a fundamental principle without putting any restrictions on portal couplings.

\subsubsection{A real pseudo-Goldstone boson, or axion-like particle (ALP),  candidate}
\label{sec:real-scalar}


Let us then consider the same potential $V(H,\phi)$ as introduced in \cref{eq:V1}. So far we have assumed that only the Higgs doublet $H$ develops a VEV  to break the EW symmetry. In a more generic approach one can also consider that the $\phi$ scalar also acquires a non-zero VEV, that we will denote as $v_\sigma$ in what follows. For a non-zero $v_\sigma$ the global $\U{G}$ symmetry gets broken and, as a result of Goldstone's theorem, a new massless real scalar emerges in the physical particle spectrum. However,  this solution serves neither the astrophysical purposes of this article nor it is cosmologically favoured. For instance, observations of the Bullet Cluster strongly disfavour models with massless scalars \cite{Bento:2000ah,Barger:2008jx,Randall:2007ph,McDonald:2007ka}. Furthermore, it is typically argued that continuous global symmetries are marginally violated by quantum gravitational effects such that they are not exact, but approximate \cite{Hui:2016ltb}. Based on these arguments we can introduce a small soft $\U{G}$ breaking mass parameter in the scalar potential of \cref{eq:V1} such that~\footnote{Note that because $\phi$ is a real field we could also add cubic terms to the potential as they would also break the symmetry softly. However, the inclusion of only dimension one and two terms is still consistent with the renormalizability of the model because these terms can only modify the minimum conditions and the propagators. This in turn means that no new infinities are generated and dimension three terms can safely be neglected.} 
\begin{equation}
   V(H,\phi) \to V(H,\phi) + V_\mathrm{soft} \qquad \textrm{with} \qquad V_\mathrm{soft} = \frac12 \mu_s^2 (\phi^2 + \phi^{\ast 2})\, .
    \label{eq:Vsoft}
\end{equation}

The Goldstone mode can be described as a phase, $\theta$, in the field space such that $\phi$ can be generically expressed as
\begin{equation}
    \phi = \frac{1}{\sqrt{2}} (\sigma + v_\sigma) e^{i \theta/v_\sigma}\,,
    \label{eq:phi}
\end{equation}
with $\sigma$ denoting the quantum fluctuations in the radial directions around the classical field configuration $v_\sigma$, such that the soft potential can be recast as
\begin{equation}
    V_\mathrm{soft} = \frac12 \mu_s^2 (v_\sigma + \sigma)^2 \cos{\(\frac{2 \theta}{v_\sigma}\)}\,.
    \label{eq:Vsoft-1}
\end{equation}
Using the field expansion in \cref{eq:phi}, the minimization conditions of the new potential in \cref{eq:Vsoft,eq:Vsoft-1} read as
\begin{equation}
\begin{aligned}
    &\mu_H^2 = -\frac12 \left( v_h^2 \lambda_H + v_\sigma^2 \lambda_{H \phi} \right) \\
    &\mu_\phi^2 = -\frac12 \left( v_\sigma^2 \lambda_\phi + v_h^2 \lambda_{H \phi} + 2 \mu_s^2 \cos{\frac{2 \theta}{v_\sigma}} \right)~\textrm{for}~\theta = n \pi v_\sigma\,, ~ n \in \mathbb{Z} \ ,
\end{aligned}
    \label{eq:tad1}
\end{equation}
such that, excluding the gauge Goldstone directions $\omega_{1,2}$ and $z$, see \cref{eq:H}, the mass matrix has the form
\begin{equation}
\bm{M}^2 =
\begin{pmatrix}
 v_h^2 \lambda_H & v_h v_\sigma \lambda_{H \phi} & 0 \\
 v_h v_\sigma \lambda_{H \phi} & v_\sigma^2 \lambda_\phi & 0 \\
0 & 0 & -2 \mu_s^2 
\end{pmatrix}\,.
\label{eq:hess2}
\end{equation} 
Note that the minimization condition along the angular direction, $\theta$, reads as
\begin{equation}
    \frac{\partial V}{\partial \theta} = -v_\sigma \mu_s^2 \sin{\frac{2 \theta}{v_\sigma}} = 0 \ ,
\end{equation}
whose solutions correspond to $\theta = n \pi v_\sigma$ as indicated in \cref{eq:tad1}. Notice that, for the $\U{}$ preserving part of the potential, the angular degrees of freedom can be \textit{gauged away} by expressing $\phi$ as in \cref{eq:phi}. On the other hand, the inclusion of explicit breaking terms implies that, in that sector, such a gauge freedom is no longer present. While at an energy-scale $\Lambda \gg \mu_s$ the angular direction is effectively flat, physically relevant effects are induced by $V_\mathrm{soft}$ and must be considered as we discuss below.

Rotating $\bm{M}^2$ to the mass basis one obtains
\begin{equation}
\bm{m}^2 = {O^\dagger}_{i}{}^{m} M_{mn}^2 O^{n}{}_{j} = 
\begin{pmatrix}
m_{h_1}^2 & 0 & 0\\ 
0   & m_{h_2}^2 & 0 \\
0 & 0 & m_\theta^2
\end{pmatrix}\,,
\end{equation}
where the eigenvalues are
\begin{equation}
\begin{aligned}
    &m_{h_{1,2}}^2 = \frac12\[v_h^2 \lambda_{H} + v_\sigma^2 \lambda_\phi \mp \sqrt{v_h^4 \lambda_H^2 + v_\sigma^4 \lambda_\phi + 2 v_h^2 v_\sigma^2 \( 2 \lambda_{H \phi}^2 - \lambda_H \lambda_\phi \)} \]\,, \\
    &m_\theta^2 = -2 \mu_s^2\,,
\end{aligned}
\label{eq:eigvals}
\end{equation}
and the orthogonal rotation matrix $\bm{O}$ reads as
\begin{equation}
\bm{O} = 
\begin{pmatrix}
\cos \alpha & \sin \alpha & 0 \\
-\sin \alpha & \cos \alpha & 0 \\
0& 0 & 1
\end{pmatrix}\,.
\label{eq:rotmat}
\end{equation}
The physical basis vectors $h_1$ and $h_2$ are written in terms of the gauge eigenbasis ones $h$ and $\sigma$ as follows:
\begin{equation}
\begin{pmatrix}
h_1 \\
h_2 \\
\theta
\end{pmatrix}
=
\bm{O}
\begin{pmatrix}
h \\
\sigma \\
\theta
\end{pmatrix}\,.
\label{eq:trans}
\end{equation}
Note that the scalar potential in \cref{eq:Vsoft-1} is a periodic function of the pseudo-Goldstone boson $\theta$. This means that, upon expansion of the cosine, only even powers of $\theta$ are allowed in a polynomial form of the mass basis scalar potential. This also means that with the soft breaking terms $\phi^2 + \phi^{\ast 2}$ the theory is invariant under a remnant discrete $\mathbb{Z}_2 \subset \U{G}$ symmetry where the pseudo-Goldstone transforms as $\theta \to - \theta$.

\begin{figure}[h!]
    \centering
    \includegraphics[width=0.25\textwidth]{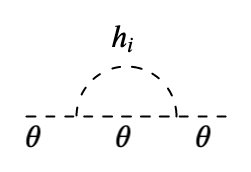}
        \includegraphics[width=0.25\textwidth]{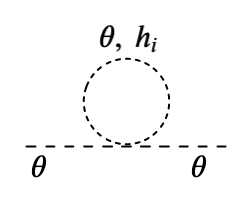}
	\caption{Leading quantum corrections to the mass of the pseudo Goldstone boson.}
	\label{fig:FT1}
\end{figure}

This model offers three real scalars. First, one of the Higgs bosons, either $h_1$ or $h_2$, must be the SM-like Higgs while the other one is a new scalar that can  be heavier or lighter than 125 GeV. 
Second, the pseudo-Goldstone boson $\theta$ receives its mass from a soft-breaking parameter and is a candidate for an ultralight real scalar. The size of its mass is a consequence of a marginal violation 
of the global $\U{G}$ which is well motivated to be induced via quantum gravitational effects - see e.g.~\cite{Harlow:2018tng}.
Before moving to the discussion on the possibility of having ultralight scalars we note that in the non-linear formulation given by \cref{eq:phi}, cubic interactions between the pseudo-Goldstone $\theta$ and the $h_{1,2}$ physical scalars appear due to the kinetic term
\begin{equation}
\begin{aligned}
    \partial_\mu \phi^* \partial^\mu \phi  &= \frac{1}{2} \left[ (\partial \sigma)^2 +  (\partial \theta)^2 \, \, \frac{{(\sigma + v_\sigma)}^2}{v_\sigma^2}  \right ]  \\
    &=   \frac{1}{2} \left[ (\partial \sigma)^2 +  (\partial \theta)^2 \right ] 
   + \frac{1}{v_\sigma}(\partial \theta)^2 \, \sigma + \frac{1}{2 v_\sigma^2}  (\partial \theta)^2 \, \sigma^2 \, .
\end{aligned}
    \label{eq:phi2}
\end{equation}
Taking the cubic coupling $(\partial \theta)^2 \, \sigma$ and using the equation of motion one can show that for assimptotically free fields the following relation holds
\begin{equation}
\begin{aligned}
    (\partial \theta)^2 \, \sigma  &= - (\partial_\mu \theta) ( \partial^\mu \sigma) \theta - \sigma \theta (\partial \cdot \partial) \theta = \frac{1}{2}   \theta^2  \, (\partial \cdot \partial) \,  \sigma + \sigma m_\theta^2 \theta^2 \\
  &= - \frac{1}{2}  \left[
  \sin \alpha \,  m_{h_1}^2 h_1 + \cos \alpha \, m_{h_2}^2 h_2
  \right] \theta^2 +  \sigma m_\theta^2 \theta^2 \, .
\end{aligned}
    \label{eq:phi3}
\end{equation}
Expanding the soft breaking potential in \cref{eq:Vsoft-1} yields a cubic coupling $\tfrac{m_\theta^2 }{v_\sigma^2}\sigma \theta^2$ which cancels the one coming from \cref{eq:phi2,eq:phi3}.
The non-linear formulation clearly shows that when dealing with low energy processes the Goldstone nature of $\theta$
is manifest since its mass is proportional to the soft breaking parameter. Radiative corrections to the mass are proportional to the self-energy diagrams shown in \cref{fig:FT1}. This formulation allows us to understand in a simple manner why corrections to the mass vanish in the limit $m_\theta \to 0$.  
In the non-linear formulation the diagram on the left has a term proportional to $m_\theta^4/v_\sigma^2$ and another term proportional to $p^4/v_\sigma^2$. Therefore,
because corrections to the mass are calculated at $p^2 = m_\theta^2$, in the limit $m_\theta \to 0$ the quantum corrections to the mass vanish and the global $\U{G}$ is recovered 
(see also discussions in~\cite{Burgess:1998ku, PhysRevLett.119.191801, Azevedo:2018exj, Glaus:2020ihj}).

In what follows $h_1$ will always denote the SM-like Higgs boson, and it can be either heavier or lighter than $h_2$. The fact that $h_2$ can be lighter than the Higgs boson raises a question on how light can it be, 
in particular, if it can become a candidate for a second ultralight real scalar. As previously discussed it is not sufficient to require a super tiny mass if quantum corrections play a dominant role. 
In the complex scalar case with no spontaneous breaking of the global symmetry via the appearance of a singlet VEV, we concluded that in order to have an ultralight scalar we would
need either a fine-tuned or a feebly coupled scenarios. Therefore in the limit $v_\sigma \to 0$ the discussion in the previous section for the unbroken scenario applies. 
On the other hand, because $v_h$ is fixed by the W boson mass, if we take the limit $v_\sigma \to \infty$, $m_{h_2} \to \infty$ the theory decouples. 

Let us now assume that we have an ultralight pseudo-Goldstone boson. 
Before moving to the phenomenological implications of this choice, we present the cubic and quartic couplings in the scalar sector that involve at least one $\theta$ field, and the coupling modifier for the Higgs couplings
to the remaining SM particles
\begin{equation}
    \begin{aligned}
    &\lambda_{_{\theta \theta h_i}} = \frac{m_{h_i}^2}{v_\sigma^2} \, \mathcal{O}_{2i}
    \\
    &\lambda_{_{\theta \theta h_i h_j}} = \frac14 \left[  \lambda_\phi\, \mathcal{O}_{2i} \mathcal{O}_{2j} + (-1)^{i+j} \lambda_{H\phi} \mathcal{O}_{1i} \mathcal{O}_{1j} \right]  \\
    &\lambda_{_{ h_i \, SM \, SM}} = \mathcal{O}_{1i} \, g_{SM}
    \end{aligned}
    \label{eq:self111}
\end{equation}
with $\mathcal{O}_{11}=\mathcal{O}_{22}= \cos \alpha$ and $\mathcal{O}_{12}= - \mathcal{O}_{21}= - \sin \alpha$. We choose as input parameters the particle masses, the rotation angle $\alpha$
and the singlet VEV $v_\sigma$. The quartic couplings can be written in terms of these parameters as follows
\begin{equation}
    \begin{aligned}
    &\lambda_{_{H\phi}} =  \frac{\sin 2\alpha \, (m_{h_1}^2 - m_{h_2}^2 )}{2\, v_\sigma \, v_h}
    \\
&\lambda_{_{\phi}} =  \frac{\cos^2 \alpha \, m_{h_2}^2 + \sin^2 \alpha \,  m_{h_1}^2}{ \, v_\sigma^2} 
  \\
&\lambda_{_{H}} =  \frac{\cos^2 \alpha \, m_{h_1}^2 + \sin^2 \alpha \,  m_{h_2}^2}{ \, v_h^2}\,.
    \end{aligned}
    \label{eq:self112}
\end{equation}
Finally, quartic self interactions of the pseudo-Goldstone come entirely from $V_\mathrm{soft}$ and read as
\begin{equation}
    \lambda_{_{\theta \theta \theta \theta}} = - \frac{m_\theta^2}{6 v_\sigma^2}\,.
    \label{eq:self113}
\end{equation}
Notice that the latter is valid in the vicinity of each of the $n \in \mathbb{Z}$ physically equivalent minima of the theory when $\theta / v_\sigma < 1$, and relevant only when $m_\theta \sim v_\sigma$. Otherwise, for $m_\theta \ll v_\sigma$, $\lambda_{_{\theta \theta \theta \theta}} \to 0$ such that $\theta$ can be treated as a free ultralight real scalar.

Let us now discuss the possible astrophysical and collider physics consequences.
First, the collider phenomenology of the model is perfectly sound~\cite{Azevedo:2020fdl, Coimbra:2013qq, Muhlleitner:2020wwk}. One of the Higgs bosons is the 125 GeV scalar, the second Higgs boson is free to be much heavier or much lighter than the SM Higgs. Althouh several searches for extra scalars have already been performed at LEP, at the Tevatron and at the LHC,
the mixing angle between the two scalars can always be made small enough that the production rates of the non-SM scalar are too small to be detected.
 The non-SM Higgs can be very light but if we move very far away from the electroweak scale radiative corrections start playing an important role. We will then be in 
    some kind of fine-tuned region as explained in detail before. The measurement of the 125 GeV Higgs couplings to the SM particles set a constraint on the mixing angle which reflects via unitarity
    in the other Higgs boson's coupling to the SM particles. At present the bound is about $|\sin \alpha| < 0.3$~\cite{ATLAS:2016neq,ATLAS:2019qdc,Robens:2021rkl}  (note that
    at the level of the cross section and branching ratios this corresponds to $\cos^2 \alpha > 0.91$). Assuming that $m_\theta \approx 0$ the partial width of a Higgs boson decaying into $\theta$ 
    particles is given by
\begin{equation}
\Gamma (h_i \to \theta \theta) = \frac{1}{32\, \pi^2} \frac{m_{h_i}^3}{v_\sigma^2} \mathcal{O}_{i1}^2 \ ,
\end{equation}
and for the SM-like Higgs $h_1$ we obtain a constraint from the measurement of the invisible Higgs decay which yields an upper bound 
on the Higgs invisible branching ratio equal to 0.11~\cite{ATLAS:2020kdi}. We can find an order of magnitude for $\mathcal{O}_{i1}^2/v_\sigma^2$
using the constraints on the $\mathcal{O}_{i1}$ obtained from the precision measurements on the Higgs couplings. Taking $|\sin \alpha| =0.3$ and the Higgs boson width to be 4 MeV
we obtain $v_\sigma > 75$ GeV. As the value of $\sin \alpha$ decreases, and its value is of course compatible with zero, the constraint on $v_\sigma$ will become increasingly weaker.

The pseudo-Goldstone boson mass is not bounded by any collider constraints. This ultralight scalar can be produced at the LHC or any other future collider via the portal coupling. As a consequence at least one of the Higgs boson has to participate in the process. This will lead to the usual final states in typical searches for dark matter that include mono-jet, mono-W, Z, and mono-Higgs events or two jets and missing energy as discussed in~\cite{Huitu:2018gbc}. In these processes the SM particle is the one identified together with missing energy.
Processes with neutrinos in the final state constitute the main background to the signal and a good modelling of both is very important.
  
In summary, the model under consideration in this section provides an extraordinary candidate for an ultralight real scalar. Its mass is always protected against quantum corrections irrespective of the scales involved in the theory, which follows from a marginal explicit violation of the global $\U{G}$ symmetry. This can simply come as a mass term or also include sizeable self-interactions. It is also worth mentioning that models with extended scalar sectors offer the possibility for strong first order phase transitions between different vacua that may have occurred in the early Universe. Such phase transitions can manifest themselves in the form of a stochastic background of primordial GWs, potentially at the reach of LISA or future experiments. We must note that constraints on ultralight scalar masses  have already been set using measurements of the spin of astrophysical BHs~\cite{cardoso2018constraining,ng2021constraints,davoudiasl2019ultralight} and searches of continuous gravitational waves~\cite{d2018semicoherent,tsukada2019first,palomba2019direct}. However, these studies assume a scalar field coupled only to gravity. Self-interactions and couplings to other particles can affect the superradiant instability and may alter the excluded mass ranges~\cite{rosa2018stimulated,Ikeda:2018nhb,bovskovic2019axionic,fukuda2020aspects}.


As discussed above, when $m_\theta \sim v_\sigma$ the quartic self couplings $\lambda_{_{\theta \theta \theta \theta}}$ can be sizeable and an astrophysical impact can be expected when compared to a non-interacting limit. Quartic self-interactions increase the maximum mass of bosonic stars and oscillatons~\cite{balakrishna1998dynamical,guzman2004evolving,escorihuela2017quasistationary}, without requiring a lower boson particle mass, and can even quench some instabilities associated with rotating stars~\cite{siemonsen2021stability,dmitriev2021instability}. To get a measure on the impact of self interactions, for the case  of scalar boson stars obtained with the model~\eqref{action} with~\eqref{LS} and $V_{(0)}^{\rm  int}(\bar \Phi \Phi)=\lambda| \bar{\Phi}\Phi| ^2$, the maximal mass of the corresponding spherical fundamental boson stars scales differently from~\eqref{mini}~\cite{Colpi:1986ye,Herdeiro:2015tia}:
\begin{equation}
 M_{\rm ADM}^{\rm max} \simeq 0.062\sqrt{\lambda} \frac{M_{\rm Pl}^3}{\mu^2}  \simeq 0.062\sqrt{\lambda}M_\odot\left(\frac{\rm GeV}{\mu}\right)^2\ .
\label{quartic}
\end{equation}
Thus,  $\lambda\neq 0$ could, in principle, increase the maximal mass to the point where astrophysical masses would not require ultralight bosons.
In the context of the model we have just described, these self-interactions can be, at most, as large as $4 \pi$. This is an interesting alternative to the ultralight particles paradigm, in the context of bosonic stars; but it is not without challenges. Constructing the gravitational solutions starting from the mini-bosonic stars ($i.e.$ without self-interactions) and then slowly increasing a dimensionless coupling $\Lambda$,  defined as
\begin{equation}
    \frac{\lambda}{4\pi}=\Lambda \frac{\mu^2}{M_{\rm Pl}^2}\simeq \frac{\Lambda}{10^{38}}\left(\frac{\mu}{\rm GeV}\right)^2 \ ,
\end{equation}
until reaching $\lambda$ of order unity is not possible. Indeed, even for fairly large values of $\Lambda\sim (\mathcal{O}(10^3))$ and even if one takes $\mu \sim \mathcal{O}(1 \ {\rm GeV})$, $\lambda$ is still extremely small. In an alternative approch one may consider the Einstein-Klein-Gordon equations in the limit of large self-interactions~\cite{Ryan:1996nk}. Conversely, the known solutions of self-interacting bosonic stars, $e.g.$ with $\Lambda\sim \mathcal{O}(10^3)$~\cite{Herdeiro:2015tia}, correspond to feeble self-interactions (an extremely small $\lambda$).

Let us also comment that considering even higher order self-interactions, for instance sextic, leads to a qualitative novelty: $Q$-balls. These are complex scalar field solitons, obtained with a non-renormalizable self-interaction, arising in some effective field theories
\cite{Coleman:1985ki,Lee:1991ax}.
However, differently from mini-boson stars (or the ones with a quartic self-interaction), this type of configurations does not trivialize in the limit of vanishing gravitational constant, $i.e.$ in the flat spacetime limit. 
$Q$-balls have a rich structure and found a variety of physically interesting applications; for example, they
appear in supersymmetric generalizations of the SM \cite{Kusenko:1997zq}, and have been suggested to generate
baryon number or to be dark matter candidates \cite{Kusenko:1997si}.
Again, they also possess 
generalizations with a rotating black hole horizon at their center \cite{Herdeiro:2014pka}.

\subsubsection{A free complex pseudo-Goldstone boson candidate}
\label{sec:free-complex}

In this section we discuss the possibility of ultralight complex scalars which are relevant for compact objects such as boson stars. As we have demonstrated above, spontaneously broken global $\U{}$ symmetries, in combination with tiny soft-breaking effects, offer a mechanism to generate extremely small masses protected against quantum effects. The same principles must then apply in order to generate ultralight complex scalars.

Let us consider an extension of the real pseudo-Goldstone model with an extra global $\U{G'}$ symmetry and a second scalar that we denote as $\phi_2$. Without loss of generality we assume that the Higgs doublet $H$ is neutral under the product group $\U{G} \times \U{G^\prime}$ but the two complex scalars $\phi_1$ and $\phi_2$ possess non-trivial Noether charges such that the theory is invariant under the transformations
\begin{equation}
    \phi_1 \to e^{i q \alpha} \phi_1 \qquad  \phi_2 \to e^{i q \alpha} \phi_2 \qquad \phi_1 \to e^{i q_1 \alpha'} \phi_1 \qquad \phi_2 \to e^{i q_2 \alpha'} \phi_2 \quad \textrm{with} \quad q_1 \neq q_2\,.
\end{equation}
The model also contains an exact discrete $\mathbb{Z}_2$ interchange symmetry that acts on $\phi_1$ and $\phi_2$ as follows
\begin{equation}
    \phi_1 \to \phi_2 \qquad \phi_2 \to \phi_1\,.
\end{equation}
Defining $\phi^i \equiv \(\phi_1,\phi_2\)$, the scalar potential, invariant under the EW and the $\U{G} \times \U{G'} \times \mathbb{Z}_2$ symmetry reads as
\begin{equation}
	V\(H,\phi_1,\phi_2\) = V_0\(H\) + \mu_\phi^2 \phi_i^\ast \phi^i + \frac{1}{2} \lambda_\phi \abs{\phi_i^\ast \phi^i}^2
	+ \lambda_{H \phi} H^\dagger H \phi^\ast_i \phi^i + \lambda_{12} \phi^\ast_1 \phi_1 \phi^\ast_2 \phi_2\,.
	\label{eq:V-complex}
\end{equation}
The continuous global symmetry $\U{G} \times \U{G'}$ can be marginally broken by the explicit mass terms
\begin{equation}
    V_\mathrm{soft} = \frac12 \mu_s^2 \( \phi_1^2 + \phi_2^2 + \mathrm{c.c.} \)
    \label{eq:soft-2}
\end{equation}
which we require to preserve the $\mathbb{Z}_2$ symmetry. Such interchange symmetry has a crucial role in allowing a solution with a complex pseudo-Goldstone physical state as we discuss below. The $\phi_1$ and $\phi_2$ fields can be generically expressed as
\begin{equation}
    \phi_1 = \frac{1}{\sqrt{2}} (\sigma_1 + v_{\sigma_1}) e^{i \theta_1/v_{\sigma_1}} \qquad \phi_2 = \frac{1}{\sqrt{2}} (\sigma_2 + v_{\sigma_2}) e^{i \theta_2/v_{\sigma_2}}\,.
    \label{eq:phi-2}
\end{equation}
The interchange symmetry also imposes that the ground state of the theory is such that $v_{\sigma_1} = v_{\sigma_2} \equiv v_\sigma$, and the minimization conditions read as
\begin{equation}
\begin{aligned}
    &\mu_H^2 = -\frac12 \left( v_h^2 \lambda_H + 2 v_\sigma^2 \lambda_{H \phi} \right) \qquad \textrm{and}
    \\
    &\mu_\phi^2 = -\frac12 \[ v_\sigma^2 \( \lambda_\phi + \lambda_{12} \) + v_h^2 \lambda_{H \phi} + 2 \mu_s^2 \cos{\frac{2 \theta_{1,2}}{v_\sigma}} \]~~\textrm{for}~~\theta_{1,2} = n \pi v_\sigma,~~n\in \mathbb{Z}\,.
\end{aligned}
    \label{eq:tad2}
\end{equation}
The $v_\sigma$ VEVs completely break the $\U{G}\times\U{G'}$ symmetry such that at the groud state the mass matrix can be written in a block diagonal form $\overline{\bm{M}}^2 = \mathrm{diag}\( \bm{M}^2, \bm{M}^2_\mathrm{PG} \)$ with
\begin{equation}
\bm{M}^2 =
\begin{pmatrix}
 v_h^2 \lambda_H & v_h v_\sigma \lambda_{H \phi} & v_h v_\sigma \lambda_{H \phi} \\
 v_h v_\sigma \lambda_{H \phi} & v_\sigma^2 \lambda_\phi & v_\sigma^2 \lambda_{12} \\
v_h v_\sigma \lambda_{H \phi} & v_\sigma^2 \lambda_{12} & v_\sigma^2 \lambda_\phi 
\end{pmatrix}
\qquad
\bm{M}^2_\mathrm{PG} =
\begin{pmatrix}
 -2 \mu_s^2 & 0 \\
 0 & -2 \mu_s^2 
\end{pmatrix}\,.
\label{eq:hess3}
\end{equation} 
We can now rotate $\bm{M}^2$ to the mass basis by means of an orthogonal transformation,
\begin{equation}
\bm{m}^2 = {O^\dagger}_{i}{}^{m} M_{mn}^2 O^{n}{}_{j} = 
\begin{pmatrix}
m_{h_1}^2 & 0 & 0\\ 
0   & m_{h_2}^2 & 0 \\
0 & 0 & m_{h_3}^2
\end{pmatrix}\,,
\end{equation}
where the rotation matrix $\bm{O}$ reads as
\begin{equation}
\bm{O} = 
\begin{pmatrix}
\cos \alpha_h & \frac{\sin \alpha_h}{\sqrt{2}} & \frac{\sin \alpha_h}{\sqrt{2}} \\
\sin \alpha_h & -\frac{\cos \alpha_h}{\sqrt{2}} & -\frac{\cos \alpha_h}{\sqrt{2}} \\
0 & -\frac{1}{\sqrt{2}} & \frac{1}{\sqrt{2}}
\end{pmatrix}\,,
\label{eq:rotmat-2}
\end{equation}
such that the physical eigenstates can be expressed as
\begin{equation}
\begin{pmatrix}
h_1 \\
h_2 \\
h_3
\end{pmatrix}
=
\bm{O}
\begin{pmatrix}
h \\
\sigma_1 \\
\sigma_2
\end{pmatrix}\,.
\label{eq:trans-2}
\end{equation}

The mass eigenvalues of the three real scalars $h_1$, $h_2$ and $h_3$, are given by
\begin{equation}
    \begin{aligned}
    &m_{h_{1,2}}^2 = \frac12 \[ v_h^2 \lambda_H + v_\sigma^2 (\lambda_\phi +\lambda_{12}) \pm \sqrt{v_h^4 \lambda_H^2 + v_\sigma^4 (\lambda_\phi + \lambda_{12})^2 - 2 v_h^2 v_\sigma^2 \(\lambda_H (\lambda_\phi + \lambda_{12}) - 4 \lambda_{H \phi}^2 \)} \]\,,
    \\
    &m_{h_{3\phantom{,2}}}^2 = v_\sigma^2 (\lambda_\phi - \lambda_{12})\,,
   \label{eq:mass123}
    \end{aligned}
\end{equation}
such that we define $h_1$ to be the SM Higgs boson while $h_2$ and $h_3$ two new real scalars. The pseudo-Goldstone mass matrix $\bm{M}^2_\mathrm{PG}$ is already diagonal yielding two degenerate real scalars with mass
\begin{equation}
    m_{\theta_1}^2 = m_{\theta_2}^2 = - 2 \mu_s^2\,.
\end{equation}
Note that such a mass equality is exact and follows from the $\mathbb{Z}_2$ interchange symmetry, which remains intact in the ground state of the theory. Furthermore, not only the masses are degenerate but also all interaction terms, which allows one to redefine the two real scalars in terms of a complex field $\eta$ and its conjugate $\eta^\ast \equiv \bar{\eta}$ as follows
\begin{equation}
    \eta = \frac{1}{\sqrt{2}} \( \theta_1 + i \theta_2 \) \qquad \bar{\eta} = \frac{1}{\sqrt{2}} \( \theta_1 - i \theta_2 \)\,,
    \label{eq:eta}
\end{equation}
with mass
\begin{equation}
    m_\eta^2 = -2\mu_s^2\,.
    \label{eq:meta}
\end{equation}
Provided that the $\U{G} \times \U{G'}$ symmetry is marginally violated by the soft parameter $\mu_s^2$, the $\eta$ field is the only ultralight complex scalar candidate that the model under discussion can offer. Even though $\eta$ is a complex field there is no Noether charge or any residual $\U{}$ symmetry associated to it. Therefore, it is important to discuss how does $\eta$ and $\bar{\eta}$ appear in a polynomial form of the scalar potential. To see this let us look at the terms involving only $\theta_1$ and $\theta_2$ which read
\begin{equation}
    V(\theta_1,\theta_2) = \frac{1}{2} v_\sigma^2 \mu_s^2 \[ \cos{\(\frac{2 \theta_1}{v_\sigma}\)} + \cos{\(\frac{2 \theta_2}{v_\sigma}\)} \]\,.
\end{equation}
Using $\eta$ and $\bar{\eta}$ as defined in \cref{eq:eta} and expanding the cosines to the fourth power of the fields, valid for $\theta_{1,2}/v_\sigma < 1$,  one obtains the following potential
\begin{equation}
    V(\theta_1,\theta_2) \to V(\eta,\bar{\eta}) = -2\mu_s^2 \bar{\eta} \eta + \frac{\mu_s^2}{v_\sigma^2} \abs{\bar{\eta} \eta}^2 + \frac{1}{6} \frac{\mu_s^2}{v_\sigma^2} \( \eta^4 + \bar{\eta}^4 \) + \cdots\,,
    \label{eq:Veta}
\end{equation}
where the last term clearly does not preserve $\U{}$ transformations. This means that for a small $\mu_s^2/v_\sigma^2$ ratio $\eta$ possesses an accidental, or approximate, Noether charge such that the ground state of the theory contains a free, ultralight, complex scalar with relevance for astrophysical observables such as boson stars. Interaction terms involving the $h_1$, $h_2$ and $h_3$ real scalars are obtained using the same procedure discussed in the previous section. Notice that the $\eta$ field is not the only ultralight candidate. However, the difference now is that instead of one single additional real scalar we can have two. For instance, a tiny new physics scale $v_\sigma \lll v_h$ can drive $h_2$ and $h_3$ to be ultralight with potential astrophysical relevance. With this in mind one can obtain the self-interactions involving only the $\eta$ fields
\begin{equation}
    \lambda_{_{\eta \eta \bar{\eta} \bar{\eta}}} = \frac{\mu_s^2}{v_\sigma^2}
    \qquad
    \lambda_{_{\eta \eta \eta \eta}} = \lambda_{_{\bar{\eta} \bar{\eta} \bar{\eta} \bar{\eta}}} = \frac{1}{6} \frac{\mu_s^2}{v_\sigma^2}\,,
    \label{eq:self_eta_U1U1}
\end{equation}
while, for completeness, non-zero quartic and cubic couplings involving $h_2$ and $h_3$ are given in \cref{app:self1}~\footnote{In this appendix we only show the trilinear scalar couplings that include at least one ultralight scalar candidate. Although in many models more than one scalar can play the role of the SM-Higgs, as it is the case for the model under discussion, when choosing the Feynman rules to be presented we opted to consider $h_1$ to be the SM-like Higgs. The complete set of Feynman rules for all models are available from the authors upon request.}.

Similarly to the case of the real pseudo-Goldstone model, the complex scalar $\eta$ possesses the same protection against quantum corrections to its mass, remaining ultralight irrespective on how strongly it couples to the SM. This model has two scalars that are candidates to be the SM-like Higgs, $h_1$ and $h_2$. The couplings of $h_1$ ($h_2$) to the remaining SM particles are just the SM Higgs couplings multiplied by $\cos \alpha_h$ ($\sin \alpha_h$). The mixing angle $\alpha_h$ is constrained from Higgs couplings measurements to the SM particles at the LHC to be $\abs{\sin \alpha_h} < 0.3$ or $\abs{\cos \alpha_h} < 0.3$, if $h_1$ or $h_2$ are the SM-like Higgs, respectively. The $h_3$ scalar is essentially singlet like and it does not couple to the remaining SM particles. It is also interesting to note that if we take for instance $h_1$ to be the SM-like Higgs, as $\cos \alpha_h \to 1$, $h_1$ decouples from the ultralight scalars but we can still test the portal coupling via the coupling to the $h_2$ scalar. Still, in the same limit, $h_2$ decouples from the SM particles and probing the portal coupling would become increasingly cumbersome at colliders. Away from this limit all particles are in principle observable at the next LHC run.

The real scalar masses $m_{h_1}^2$, $m_{h_2}^2$ and $m_{h_3}^2$, if ultralight (note that one of scalars $h_1$ or $h_2$ is the SM-like Higgs with a mass of 125 GeV), may receive enormous quantum corrections from Higgs boson loops unless the quartic couplings $\lambda_{_{ h_i h_j h_k h_l}}$, as well a the trilinear couplings $\lambda_{_{ h_i h_j h_k}}$, see \cref{app:self1}, are also extremely small. In practice, this corresponds to an equally tiny portal coupling $\lambda_{H \phi}$ essentially closing the portal between the ultralight sector and the SM particles. Therefore, if we expect the couplings to be perturbative but not too small for the portal to be closed, the masses of $h_1$ and $h_2$ should be of the order of the electroweak scale. The mass of $h_3$ is only proportional to the singlet VEV and so it lives, to a large extent, in a different mass scale. Note however that when  $v_\sigma \to 0$ one of the $h_1$ or $h_2$ scalars will be the 125 GeV while the other together with $h_3$ will become massless. 


To finalize this section, we discuss potential astrophysical and collider physics implications. A model with an ultralight sector containing one complex and two real scalars offers a rather rich astrophysical phenomenology. The two real states can leave characteristic GW signatures when interacting with astrophysical BHs, as explained above for the real case.
 On the other hand, the complex scalar field can form stable boson stars and clouds in equilibrium with black holes (or black hole hair); such boson stars or hairy black holes, in binaries, would merge producing GWs that could be detected by current observations, could provide a smoking gun signature for the bosonic field and thus the bosonic particle. A proof of concept is in fact given by~\cite{CalderonBustillo:2020srq}. However, if we want to preserve the stability of the self-gravitating scalar condensate (either in black holes or in a solitonic star) one has to require a certain hierarchy between $\mu_s^2$ and $v_\sigma^2$ in order to suppress the $\U{}$ violating self interactions in \cref{eq:Veta}. Whereas quantifying this hierarchy requires a detailed study of the decay timescale of $e.g.$ oscillations obtained from the model~\eqref{eq:Veta}, at least in the spherical case, such timescale may be much larger than the Hubble time, $cf.$~\cite{Page:2003rd}, even for $\frac{\mu_s^2}{v_\sigma^2}\sim \mathcal{O}(1)$.
In such case, however, all self-interaction terms among ultralight states are sizeable and may need to be considered. However, if $v_\sigma \gg \mu_s$ the only ultralight particle is $\eta$ which can be treated as a free complex scalar.
    
In essence, the current model offers an extraordinary candidate for a free ultralight complex scalar provided that the $v_\sigma$ scale is sufficiently larger than the soft breaking parameter $\mu_s$. Depending on the model parameters, two new real scalars can also have extra astrophysical, collider or even cosmological implications. Finally, with a rich vacuum structure, there is a possibility for strong first order phase transitions having occurred in the early Universe potentially leaving a stochastic background of primordial GWs at the reach of LISA or future gravitational interferometers.

\subsubsection{A complex pseudo-Goldstone boson candidate with self-interactions}
\label{subsec:PG-complex}

The discussion of the previous model revealed that in the limit where quartic self-interactions are tiny the complex scalar acquires an accidental Noether charge. However, when such terms become relevant this is no longer the case and can affect the stability of compact objects composed (partly or in the whole) by a condensate of these scalar particles. Moreover, self-interactions can change the maximal mass of bosonic stars, as discussed in section~\ref{sec:real-scalar}.  The aim now is to propose a framework based on the same principles but allowing for self interacting complex scalars and an exact, rather then accidental, Noether charge. A possible way that one might think of consists in a modification of the previous case where we promote the global abelian $\U{G} \times \U{G'} \times \mathbb{Z}_2$ symmetry to a non-abelian one $\SU{2}{G} \times \U{G}$. In this case, instead of two scalar fields $\phi_1$ and $\phi_2$ related by an interchange symmetry, we unify them in a single $\SU{2}{G}$ doublet representation $\Phi = \( \phi_1,\phi_2 \)^\top$, such that the scalar potential can be written as
\begin{equation}
V\(H,\Phi\) = V_0\(H\) + \mu_\Phi^2 \Phi^\dagger \Phi + \frac{1}{2} \lambda_\Phi \abs{\Phi^\dagger \Phi}^2 + \lambda_{H \Phi} H^\dagger H \Phi^\dagger \Phi\,.
\label{eq:V2}
\end{equation}
Notice that, similarly to the breaking of the EW symmetry in the SM, a VEV in $\Phi$ reduces $\SU{2}{G} \times \U{G}$ down to $\U{G'}$. However, the key difference now is that we are dealing with global symmetries, and, as a result of the breaking, three Goldstone bosons, $\theta_{1,2,3}$, emerge in the physical spectrum. Recalling that Goldstone modes correspond to angular directions in the field space one can express the doublet $\Phi$ in an exponential form as
\begin{equation}
\begin{aligned}
\Phi = \dfrac{1}{\sqrt{2}} 
\begin{pmatrix}
0  \\
v_\varphi + \varphi
\end{pmatrix}
e^{\frac{i}{2} \tfrac{\tau^a \theta_a}{v_\varphi}}\,,
\end{aligned}
\label{eq:Phi}
\end{equation}
with $\varphi$ representing radial quantum fluctuations around the classical field configuration $v_\varphi$ and $\tau^a$ the three Pauli matrices. Under the assumption that the $\SU{2}{G} \times \U{G}$ symmetry is marginally violated by tiny effects from, e.g.~Quantum Gravity or other unspecified source, one can add a soft breaking potential such that the angular modes become marginally massive. Before writing it let us note a few relevant details:
\begin{enumerate}
    \item In order to preserve a Noether charge one cannot write mass terms of the form $\mu_s^2 \phi_1^2 + \mathrm{c.c.}$ where we are casting $\Phi = \( \phi_1,\phi_2 \)^\top$. To see this, if one expands \cref{eq:Phi} and define 
    \begin{equation}
        \eta^+ = \frac{1}{\sqrt{2}} \(i \theta_1 + \theta_2  \) \qquad \textrm{and} \qquad \eta^- = \frac{1}{\sqrt{2}} \(-i \theta_1 + \theta_2  \)\,,
        \label{eq:eta-2}
    \end{equation}
    the upper component of $\Phi$ and its conjugate take the form
    \begin{equation}
        \phi_1 = \eta^+ (v_\varphi + \varphi) \frac{\sin{\(\frac{\sqrt{\theta_3^2 + 2 \eta^+ \eta^-}}{v_\varphi}\)} }{\sqrt{\theta_3^2 + 2 \eta^+ \eta^-}} \qquad \textrm{and} \qquad
        \phi_1^\ast = \eta^- (v_\varphi + \varphi) \frac{\sin{\(\frac{\sqrt{\theta_3^2 + 2 \eta^+ \eta^-}}{v_\varphi}\)} }{\sqrt{\theta_3^2 + 2 \eta^+ \eta^-}}\,,
        \label{eq:phi1-expl}
    \end{equation}
    with the $+$ and $-$ superscripts denoting the sign of the global Noether charge. It is then clear that writing $\phi_1^2 + \mathrm{c.c.}$ results in the unwanted $\U{G'}$ violating operators $\eta^+ \eta^+$, $\eta^- \eta^-$ as well as higher order combinations of these operators upon power series expansion of the sine and the square roots. However, since we want to preserve a global Noether charge, we can instead define the $\SU{2}{G}$ soft breaking term $\mu_{1}^2 \phi_1^\ast \phi_1$.
    \item The $\phi_2$ component is neutral under $\U{G'}$ and can be cast as
    \begin{equation}
        \phi_2 = \frac{1}{\sqrt{2}}(v_\varphi + \varphi) \frac{\sqrt{\theta_3^2 + 2 \eta^+ \eta^-} \cos{\(\frac{\sqrt{\theta_3^2 + 2 \eta^+ \eta^-}}{v_\varphi}\)} - i \theta_3 \sin{\(\frac{\sqrt{\theta_3^2 + 2 \eta^+ \eta^-}}{v_\varphi}\)} }{\sqrt{\theta_3^2 + 2 \eta^+ \eta^-}} \ .
        \label{eq:phi2-expl}
    \end{equation}
    { Therefore, soft terms of the form $\mu_{2}^2 (\phi_2^2 + \mathrm{c.c.})$ conserve a remnant $\U{G'}$ Noether charge and can safely be added to the Lagrangian.} Furthermore, bilinear operators $\eta^+ \eta^-$ and $\theta_3^2$ become allowed generating the needed mass terms for the pseudo-Goldstone bosons. Note that for $\mu_{1}^2 = \mu_{2}^2$ the $\SU{2}{G}$ symmetry is restored and the complex fields $\eta^\pm$ become massless as we show below.
\end{enumerate}
Taking into account the two points above, the $\SU{2}{G} \times \U{G}$ symmetry can be softly broken by
\begin{equation}
   V_\mathrm{soft} = \mu_1^2 \phi_1^\ast \phi_1 + \frac12 \mu_{2}^2 \( \phi_2^2 + \mathrm{c.c.}\) \ ,
    \label{eq:VsoftSU2}
\end{equation}
such that a residual $\U{G'}$ symmetry is preserved. The soft breaking terms can be explicitly written in terms of the pseudo-Goldstone bosons as
\begin{equation}
   V_\mathrm{soft} = \frac12 \(v_\varphi + \varphi\)^2 \frac{(\mu_1^2 + \mu_2^2) \eta^+ \eta^- + \[ \mu_2^2 \theta_3^2 - \( \mu_1^2 - \mu_2^2 \) \eta^+ \eta^- \] \cos{\(\frac{2\sqrt{\theta_3^2 + 2 \eta^+ \eta^-}}{v_\varphi}\)} }{\theta_3^2 + 2 \eta^+ \eta^-}\,.
    \label{eq:Vthetas}    
\end{equation}

The minimization conditions of the full $V+V_\mathrm{soft}$ scalar potential are analogous but not as trivial as those in \cref{eq:tad1,eq:tad2}. In particular, we have
\begin{equation}
    \begin{aligned}
        &\mu_H^2 = -\frac12 (v_h^2 \lambda_H + v_\varphi^2 \lambda_{H\phi}) \\
        &\mu_\Phi^2 = -\frac12 \[ v_h^2 \lambda_{H \Phi} + v_\varphi^2 \lambda_\Phi + \frac{2 \(\mu_1^2 + \mu_2^2 \) \eta^+ \eta^- + \[2 \mu_2^2 \theta_3^2 - 2 (\mu_1^2 - \mu_2^2) \eta^+ \eta^- \] \cos\(\dfrac{2 \sqrt{\theta_3^2 + 2 \eta^+ \eta^-}}{v_\varphi}\) }{\theta_3^2 + 2 \eta^+ \eta^-} \]
        \\
        &\forall \quad \theta_3 = \sqrt{n^2 \pi^2 v_\varphi^2 - 2 \eta^+ \eta^-} \qquad \textrm{with} \qquad n \in \mathbb{Z}\,,
    \end{aligned}
\end{equation}
such that, for any $n$, the second minimization condition recovers the canonical form
\begin{equation}
    \mu_\Phi^2 = -\frac12 \(v_h^2 \lambda_{H\Phi} + v_\varphi^2 \lambda_\Phi + 2 \mu_2^2 \)\,.
\end{equation}

If we consider the limit of small field values we can expand \cref{eq:Vthetas} to the fourth power on the pseudo-Goldstone modes approximating it to the following polynomial potential
\begin{equation}
\begin{aligned}
    V_\mathrm{soft} \approx& - \mu_2^2 \theta_3^2 + \(\mu_1^2 - \mu_2^2\) \eta^+ \eta^- + \frac23\frac{\mu_2^2 - \mu_1^2}{v_\varphi^2} \abs{\eta^+ \eta^-}^2 +
  \frac13 \frac{\mu_2^2}{v_\varphi^2} \theta_3^4 + \frac{3\mu_2^2 - \mu_1^2}{3 v_\varphi^2} \eta^+ \eta^- \theta_3^2 
\end{aligned}
    \label{eq:Vthetas-2}
\end{equation}
where, contrary to \cref{eq:Veta}, it contains an exact Noether charge. Note that in the limit of $\mu_1^2 = \mu_2^2$ the complex scalar $\eta^\pm$ becomes massless and the $\SU{2}{G}$ symmetry is restored. The current model also offers a real scalar with mass 
\begin{equation}
    m_{\theta_3}^2 = -2 \mu_2^2\,,
    \label{eq:mtheta3}
\end{equation}
different than that of its complex partner
\begin{equation}
    m_{\eta^\pm}^2 = \mu_1^2 - \mu_2^2\,.
    \label{eq:metapm}
\end{equation}
Note that the model contains enough freedom to allow a mass hierarchy between the real and the complex scalars such that only one of them can be chosen to be of astrophysical relevance, therefore agreeing with current excluded mass bounds for real ultralight scalar bosons. The existence of a real field is constrained by current bounds on ultralight bosonic particles. The freedom in the values of the masses makes it possible to avoid those excluded regions.  Furthermore, if $v_\varphi^2 \gg \mu_i^2$ both $\theta_3$ and $\eta^\pm$ can be treated as free particles such that the mass of the lighter does not receive quantum corrections from the heavier pseudo-Goldstone.

The model also contains two additional real scalars whose masses read as
\begin{equation}
m_{h_{1,2}}^2 = \frac12\[v_h^2 \lambda_{H} + v_\varphi^2 \lambda_\Phi \mp \sqrt{v_h^4 \lambda_H^2 + v_\varphi^4 \lambda_\Phi + 2 v_h^2 v_\varphi^2 \( 2 \lambda_{H \Phi}^2 - \lambda_H \lambda_\Phi \)} \]\,,
\label{eq:eigvals2}
\end{equation}
where we define $h_1$ to be the SM Higgs boson while $h_2$ a new real scalar. These can be written in terms of the gauge eigenbasis vectors $h$ and $\varphi$ as follows:
\begin{equation}
\begin{pmatrix}
h_1 \\
h_2
\end{pmatrix}
=
\bm{O}
\begin{pmatrix}
h \\
\varphi
\end{pmatrix}\,,
\label{eq:trans-3}
\end{equation}
where the rotation matrix reads as
\begin{equation}
\bm{O} = 
\begin{pmatrix}
\cos \alpha & \sin \alpha \\
-\sin \alpha & \cos \alpha
\end{pmatrix}\,.
\label{eq:rotmat3}
\end{equation}
We can then write the relevant self interactions involving ultralight scalars for the model under consideration. Notice that the same features discussed for $h_2$ in the two previous sections also apply here. The quartic couplings involving only pseudo-Goldstone bosons can be read from the last three terms in \cref{eq:Vthetas-2} whereas the remaining ones are given as 
\begin{equation}
    \begin{aligned}
    &\lambda_{_{\eta^+ \eta^- h_1 h_1}} = \lambda_{_{\theta_3 \theta_3 h_1 h_1}} = \frac14 \( \lambda_{H \Phi} \cos^2 \alpha + \lambda_\Phi \sin^2 \alpha \)
    \\
    &\lambda_{_{\eta^+ \eta^- h_2 h_2}} = \lambda_{_{\theta_3 \theta_3 h_2 h_2}} = \frac14 \( \lambda_{H \Phi} \sin^2 \alpha + \lambda_\Phi \cos^2 \alpha \)
    \\
    &\lambda_{_{h_1 h_1 h_2 h_2}} = \frac{1}{32} \[ 3 \lambda_H + 2 \lambda_{H \Phi} +3 \lambda_\Phi - 3 \( \lambda_H -2 \lambda_{H \Phi} + \lambda_\Phi \) \cos (4 \alpha) \]
    \\
    &\lambda_{_{h_2 h_2 h_2 h_2}} = \frac18 \( \lambda_\Phi \cos^4 \alpha + 2 \lambda_{H \Phi} \cos^2 \alpha \sin^2 \alpha + \lambda_H \sin^4 \alpha \)\,.
    \end{aligned}
    \label{eq:self-4-eta}
\end{equation}
Finally, the cubic interactions read as
\begin{equation}
    \begin{aligned}
     &\lambda_{_{\eta^+ \eta^- h_1}} = \lambda_{_{\theta_3 \theta_3 h_1}}  = \frac12 \( v_h \lambda_{H\Phi} \cos \alpha + v_\varphi \lambda_\Phi \sin \alpha \) { = \frac12 \frac{m_{h_1}^2 }{v_\varphi} \sin \alpha}
    \\
    &\lambda_{_{\eta^+ \eta^- h_2}} = \lambda_{_{\theta_3 \theta_3 h_2}}  = \frac12 \( v_\varphi \lambda_\Phi \cos \alpha - v_h \lambda_{H \Phi} \sin \alpha \) { = \frac12 \frac{m_{h_2}^2 }{v_\varphi} \cos \alpha}
    \\
    &\lambda_{_{h_1 h_2 h_2}} = \frac18 \[ v_h (3 \lambda_H + \lambda_{H \Phi}) \cos \alpha + 3 v_h (\lambda_{H\Phi} - \lambda_H) \cos(3 \alpha) + v_\varphi (\lambda_{H \Phi} + 3 \lambda_\Phi ) \sin \alpha \right. \\
   & \qquad \qquad \left. 
   + 3 v_\varphi (\lambda_\Phi - \lambda_{H \Phi}) \sin(3 \alpha) \] = \frac14 \frac{m_{h_1}^2 + 2 m_{h_2}^2}{v_h v_\varphi} (v_h \cos \alpha + v_\varphi \sin \alpha) \sin(2 \alpha),
    \end{aligned}
    \label{eq:self-3-eta}
\end{equation}
with the Lagrangian basis quartic couplings written in terms of the physical parameters as in \cref{eq:self112} identifying $\lambda_\phi$ and $\lambda_{H \phi}$ with $\lambda_\Phi$ and $\lambda_{H \Phi}$ respectively. Let us comment that larger global symmetries, as e.g.~those described by generic $\SU{N}{}$ groups, can offer a larger multiplicity of both real and complex ultralight pseudo-Goldstone modes when spontaneous and explicit symmetry breaking takes place simultaneously. 
As a final remark, the model presented in this section, in particular the details of a softly broken $\SU{2}{}$ global symmetry is, to the best of our knowledge, so far lacking in the literature and discussed here for the first time.

\subsection{Ultralight Proca fields: the case of gauge theories with spontaneous symmetry breaking}
\label{sec:proca}

In \cref{sec:soft} we have studied the case of emergent ultralight scalars in the form of pseudo-Goldstone bosons. The key feature is that a marginal violation of a continuous global symmetry generates ultralight scalar masses on the Goldstone directions, protected against quantum corrections by the underlying, approximate, symmetry. In what follows we change our paradigm and instead of global invariance under a certain transformation group we consider local or gauge symmetries. This simple modification has profound effects. In spontaneously broken gauge symmetries, the Goldstone bosons, which are non-physical, are absorbed by longitudinal modes of vector bosons and, instead of massless scalars one obtains massive spin-1 bosons. Therefore, as long as a continuous symmetry is local, there is no longer the need to invoke any explicit breaking. Also relevant is the fact that a gauge symmetry protects vector bosons from acquiring large quantum corrections.

\subsubsection{A real ultralight Proca field}
\label{sec:local-1}

Let us consider the model discussed in \cref{sec:real-scalar} but with local $\U{H}$ transformations instead of global ones. Here, H is used to denote a \textit{hidden} gauge symmetry in order to distinguish it from the ordinary $\U{Y}$ or the global $\U{G}$ so far discussed. Since gauge symmetries are well known to be exact in nature, as e.g.~$\SU{3}{C}$, that describes quantum chromodynamics, or $\U{e.m.}$, that describes the electromagnetic theory, it is no longer necessary to softly break it. Therefore, for our purposes in the current discussion, it is sufficient to consider the scalar potential in \cref{eq:V1}, which is in fact the most generic one. The scalar sector has the same properties discussed in \cref{sec:real-scalar} apart from the absence of the pseudo-Goldstone mode $\theta$. In particular, there will be two physical Higgs bosons with a mass spectrum and a single scalar mixing angle.

Let us then study the gauge sector by considering the following kinetic terms
\begin{equation}
\mathcal{L}_\mathrm{kin} \supset \frac{1}{4} B'_{\mu \nu} B'^{\mu \nu} + D_\mu \phi^\ast D^\mu \phi \qquad \text{with} \qquad D_\mu = \partial_\mu + i g'_1 B'_\mu \qquad \textrm{and} \qquad B'_{\mu \nu} = \del_\mu B'_\nu - \del_\nu B'_\mu\,.
\label{eq:Lkin2}
\end{equation}
Note that, in general a kinetic mixing term of the form $\tfrac{1}{2} \kappa B_{\mu \nu} B'^{\mu \nu}$ is also allowed resulting in a mixture of the new $B'_\mu$ gauge boson with the photon. However, there are strong phenomenological constraints on the value of $\kappa$ which must be rather small. We refer to the discussion in Sec.~B of \cite{Morais:2019aqz} where it is shown that for small kinetic mixing we can, to a good approximation, set $\kappa \to 0$. Furthermore, it is also possible to impose invariance of the Lagrangian under a discrete (dark) symmetry such that $B' \to - B'$ and $\phi \to \phi^*$, while all other particles
are even under this transformation.  

It promptly follows from the $D_\mu \phi^\ast D^\mu \phi$ term that, in the ground state $\mean{\phi} = \tfrac{1}{\sqrt{2}} v_\sigma$, the theory contains a new real gauge boson $\mathcal{B}_\mu$ with mass
\begin{equation}
    m_\mathcal{B}^2 = \frac{1}{4} {g'_1}^2 v_\sigma^2\,.
    \label{eq:mB}
\end{equation}
%
Writing  $v_\sigma = \epsilon v_h$,   in the limit where either $\epsilon$ or $g'_1$ are in the range $[10^{-31},10^{-21}]$ then $\mathcal{B}_\mu$ becomes an excellent candidate for an ultralight real Proca field with a mass in the range $10^{-20} \lesssim m_\mathcal{B} / \mathrm{eV} \lesssim 10^{-10}$. While for small $\epsilon$ we are attributing the size of the Proca field's mass to a new energy-scale well below the EW one, a tiny $g'_1$ would result in a feebly interacting theory. It is also possible that the smallness of the $\mathcal{B}_\mu$ mass results from an hybrid scenario where $10^{-31} \lesssim g'_1 \epsilon \lesssim 10^{-21}$. While a small $g'_1$ only affects the mass and coupling of the new gauge boson
a small vacuum expectation value of the singlet leads us once more to fine tuning in the scalar sector. In fact, a tiny $v_\sigma$ is obtained via $\lambda_\phi \approx \lambda_H$~\cite{Duch:2015jta} and leads to a light mass of one of the Higgs bosons. 

The fact we are discussing an abelian symmetry, means that  there are no self-interactions solely involving $\mathcal{B}_\mu$. However, the new gauge field does interact with the physical Higgs bosons $h_1$ and $h_2$ allowing the following quartic and cubic self interactions
\begin{equation}
\begin{aligned}
    &g_{_{\mathcal{B} \mathcal{B} h_i h_j}} =  2 {g'_1}^2 \mathcal{O}_{2i} \mathcal{O}_{2j} 
    \qquad
     g_{_{\mathcal{B} \mathcal{B} h_i}} = 2 m_\mathcal{B} {g'_1} \mathcal{O}_{2i} \, ,
\end{aligned}
    \label{eq:self-proca-B}
\end{equation}
with the rotation matrices given in  \cref{eq:rotmat}.
Since we are postulating a very light gauge boson, the cubic vertex with both Higgs bosons is strongly suppressed and as one would expect there are no constraints from the Higgs invisible decays. The quartic couplings can only be large
if we would be considering the fine-tuned scenario where ${g'_1}$ is not constrained while $\epsilon \to 0$.  Again considering that $h_1$ is the SM-like Higgs boson, the only quartic interaction 
that would survive $\epsilon \to 0$ would be $ g_{_{\mathcal{B} \mathcal{B} h_2 h_2}}$ which incidentally would be the interaction between four massless particles (this is because $\sin \alpha$ is also
of order $\epsilon$ in the limit $\epsilon \to 0$).
It is interesting to note that for the case of an extremely small $\U{H}$ breaking scale and a large gauge coupling $g'_1$, the relic abundance of the Proca field $\mathcal{B}$ in today's Universe 
depends on the annihilation channel $h_2 h_2 \to \mathcal{B} \mathcal{B}$ (in the $\epsilon \to 0$ scenario $h_2$ can also be part of a dark sector of the Universe)

In essence, the model under discussion can potentially offer both astrophysical and cosmological observables which may shed light on a dark sector of the Universe. Last but not least, 
the non-SM Higgs, $h_2$ can be detected at the LHC provided that the $\U{H}$ breaking scale
is of the order or larger than the EW one.

\subsubsection{A complex ultralight Proca field with self interactions}
\label{sec:local-2}

In \cref{subsec:PG-complex} we have studied the possibility for ultralight complex scalars with relevance for boson stars. Our aim now is to discuss a simple model with emergent ultralight complex vector bosons, necessary to describe the hypothesis of stable Proca stars.

A complex Proca field must be invariant under a $\U{}$ phase transformation which, for the case of gauge theories, can be regarded as a candidate for hidden electromagnetism. In fact, the same principles applied for the SM electroweak sector can be employed here in such a way that new hidden, ultralight, vector bosons emerge. In particular, complex Proca fields can be formally regarded in the same footing as the well known SM $W^\pm$ bosons.

Let us then consider an extension of the SM bosonic sector with a \textit{mirror}, or \textit{hidden}, $\SU{2}{H} \times \U{H}$ gauge symmetry. Similar concepts but at higher scales were previously discussed both in terms of astrophysical observables \cite{Curtin:2019ngc,Ciancarella:2020msu} and dark matter models \cite{Ritter:2021hgu,Kaganovich:2021jke,Foot:2004pa,Foot:2006ru}. This means that, besides a complete copy of the vector boson content, one also has a new complex scalar doublet that can be cast as in \cref{eq:Phi}. With such a formulation the communication between the mirror and the visible sectors can be realized via the scalar quartic portal coupling $\lambda_{H \Phi}$ in \cref{eq:V2} and a kinetic mixing between $\U{Y}$ and $\U{H}$. However, strong restrictions on the kinetic mixing constrain it to be rather small \cite{Merkel:2014avp,NA482:2015wmo,BaBar:2014zli,BaBar:2017tiz} such that we can safely neglect it in the remainder of the discussion.

The gauge sector can then be described by the following Lagrangian
\begin{equation}
\mathcal{L}_\mathrm{kin} \supset \frac{1}{4} B'_{\mu \nu} B'^{\mu \nu} 
+ \frac{1}{4} {F'}^a_{\mu \nu} {F'}_a^{\mu \nu}
+ D_\mu \Phi^\ast D^\mu \Phi\,,
\label{eq:Lkin3}
\end{equation}
where the gauge covariant derivative reads as
\begin{equation}
D_\mu = \partial_\mu \mathbb{1} + i g'_1 B'_\mu \mathbb{1} + i g'_2 \frac{\tau_a}{2} {A'}_\mu^a\,,
\label{eq:covD}
\end{equation}
the field strength tensors are given by
\begin{equation}
    B'_{\mu \nu} = \del_\mu B'_\nu - \del_\nu B'_\mu \qquad \textrm{and} \qquad {F'}^a_{\mu \nu} = \del_\mu {A'}_\mu^a - \del_\nu {A'}_\mu^a - g_2 \varepsilon^a_{\phantom{a} bc} {A'}_\mu^b {A'}_\nu^c\,
\end{equation}
and where $\varepsilon_{abc}$ denotes the Levi-Civita symbol while $g'_1$ and $g'_2$ are the $\U{H}$ and $\SU{2}{H}$ gauge couplings respectively. The scalar potential is identical to that in \cref{eq:V2} where the mirror doublet $\Phi$ can be expanded as in \cref{eq:Phi}.

The ground state of the mirror sector is characterized by the vacuum state
\begin{equation}
	\mean{\Phi} = \dfrac{1}{\sqrt{2}} 
	\begin{pmatrix}
	0  \\
	v_\varphi
	\end{pmatrix}\,,
\end{equation} 
which reduces the mirror symmetry according to the well known pattern $\SU{2}{H} \times \U{H} \to \U{h.e.m}$, with $\textrm{h.e.m}$ denoting \textit{hidden electromagnetism}. The scalar sector is identical to that discussed for the global $\SU{2}{G} \times \U{G}$ model but no longer containing the $\theta_3$ and $\eta^\pm$ pseudo-Goldstone state since the breaking of our gauge symmetry is purely spontaneous. In particular, the mass spectrum can be read from \cref{eq:eigvals2} while quartic and cubic self-interactions are given in the last two lines of \cref{eq:self-4-eta} and the last line in \cref{eq:self-3-eta} respectively.

It follows from the scalar kinetic terms evaluated at the vacuum, i.e.~$\mean{D_\mu\Phi^\ast D^\mu \Phi}$, that the vector bosons mass matrix reads as
\begin{equation}
\bm{M_V}^2 = v_\varphi^2
\begin{pmatrix}
\frac14 g'_2  & 0 & 0 & 0 \\ 
0 & \frac14 g'_2  & 0 & 0 \\
0 & 0 & \frac14 g'_2 & -\frac12 g'_1 g'_2  \\
0 & 0 & -\frac12 g'_1 g'_2 & \frac14 {g'}_1^2 
\end{pmatrix}\,.
\label{eq:hessGauge}
\end{equation}
We can now rotate $\bm{M_V}^2$ to the $\U{h.e.m}$ charge and mass proper basis with the following transformation
\begin{equation}
\bm{m_\mathcal{V}}^2 = {O_\mathcal{V}^\dagger}_{i}{}^{m} M_{mn}^2 O_\mathcal{V}^{n}{}_{j} =
\begin{pmatrix}
m_{\mathcal{A}^+}^2 & 0 & 0 & 0 \\
0 & m_{\mathcal{A}^-}^2 & 0 & 0 \\
0 & 0 & m_{\mathcal{B}}^2 & 0 \\ 
0 & 0 & 0  & m_{\gamma'}^2
\end{pmatrix}\,,
\end{equation}
with 
\begin{equation}
O_\mathcal{V} =
\begin{pmatrix}
\tfrac{1}{\sqrt{2}}  & \tfrac{i}{\sqrt{2}} & 0 & 0 \\ 
\frac{1}{\sqrt{2}} & -\tfrac{i}{\sqrt{2}}  & 0 & 0 \\
0 & 0 & \sin \theta'_{_W} & \cos \theta'_{_W}  \\
0 & 0 & -\cos \theta'_{_W} & \sin \theta'_{_W} 
\end{pmatrix}\, ,
\label{eq:hessGauge-1}
\end{equation}
such that the physical eigenvectors are defined as
\begin{equation}
\begin{pmatrix}
\mathcal{A}^+_\mu \\
\mathcal{A}^-_\mu \\
\gamma'_\mu \\
\mathcal{B}_\mu
\end{pmatrix}
=
\bm{O}_\mathcal{V}
\begin{pmatrix}
{A'}_\mu^1 \\
{A'}_\mu^2 \\
{A'}^3_\mu \\
B'_\mu
\end{pmatrix}\,.
\label{eq:trans-4}
\end{equation}
The gauge mixing angle $\theta'_{_W}$ is the mirror analogous of the Weinberg angle and is related to the mirror gauge couplings $g'_{1,2}$ and the hidden charge $e'$ as
\begin{equation}
    g'_1 = \frac{e'}{\cos \theta'_{_W}} \qquad \textrm{and} \qquad{} g'_2 = \frac{e'}{\sin \theta'_{_W}}\,.
\end{equation}
With the definitions above we can, at last, write the masses of the mirror gauge bosons as
\begin{equation}
	m_{\gamma'} = 0\,, \qquad m_{\mathcal{B}}^2 = \frac12 v_\varphi^2 {e'}^2 \csc^2\(2 \theta'_{_W}\) \,, \qquad m_{\mathcal{A}^\pm}^2 = \frac14 v_\varphi^2 {e'}^2 \csc^2 \theta'_{_W}\,,
	\label{eq:mirror-mass}
\end{equation}
where, for the case of a new extremely small energy scale, the model predicts a new complex vector field $\mathcal{A}^\pm_\mu$ with astrophysical relevance in the context of Proca stars. In addition, the model also offers a real Proca field $\mathcal{B}_\mu$ as well as a new, massless, hidden photon $\gamma'_\mu$. Note that the smallness of both the real and complex Proca fields can also be attributed to a feebly interacting theory where, instead of a new tiny scale $v_\varphi$, it is the value of $e'$ that sets the size of both $m_{\mathcal{A}^\pm}$ and $m_\mathcal{B}$.

The presence of a hidden photon can also have interesting phenomenological implications. In particular, it was recently proposed in \cite{Fabbrichesi:2017vma} that rare Kaon decays such as $K^+ \to \pi^+ \pi^0 \gamma'$, are sensitive channels to probe massless hidden photons where the typically searched kinetic-mixing interactions are nonviable. Furthermore, as it is discussed in \cite{Zhang:2018fbm} traces of hidden and ordinary photon mixing in the weak and electromagnetic interactions are only possible if new particles beyond those of the SM are involved, opening up the possibility for complementary searches for new physics.

To finalize this section let us write down the self-interactions involving ultralight Proca fields, the hidden photon and scalars. First, let us consider the pure gauge sector where, due to non-trivial Lorentz index contractions we explicitly write the cubic and quartic interactions as
\begin{equation}
\begin{aligned}
    \mathcal{L}_3 &= i e' \cot \theta'_{_W} \del^\nu g^{\mu \rho} \( \mathcal{B}_\rho \mathcal{A}^-_\mu \mathcal{A}^+_\nu - \mathcal{B}_\rho \mathcal{A}^+_\mu \mathcal{A}^-_\nu\) + i e' \theta'_{_W} \del^\nu g^{\mu \rho} \( \gamma'_\rho \mathcal{A}^-_\mu \mathcal{A}^+_\nu - \gamma'_\rho \mathcal{A}^+_\mu \mathcal{A}^-_\nu\)
    \\
    &+ \mathrm{Perm}(\mu,\nu,\rho)
\end{aligned}
\label{eq:L3}
\end{equation}
and
\begin{equation}
\begin{aligned}
    \mathcal{L}_4 =& -\frac14 {e'}^2 g^{\mu \rho} g^{\nu \lambda} \[ \cot^2\theta'_{_W} \( \mathcal{B}_\mu \mathcal{B}_\rho \mathcal{A}^-_\nu \mathcal{A}^+_\lambda - \mathcal{B}_\mu \mathcal{B}_\lambda \mathcal{A}^-_\nu \mathcal{A}^+_\rho\) + \( \gamma'_\mu \gamma'_\rho \mathcal{A}^-_\nu \mathcal{A}^+_\lambda - \gamma'_\mu \gamma'_\lambda \mathcal{A}^-_\nu \mathcal{A}^+_\rho\) 
    \right.
    \\
    &-
    \left.
    \cot \theta'_{_W} \( \mathcal{B}_\mu \gamma'_\rho \mathcal{A}^-_\nu \mathcal{A}^+_\lambda - \mathcal{B}_\mu \gamma'_\lambda \mathcal{A}^-_\nu \mathcal{A}^+_\rho\) +\csc^2 \theta'_{_W} \( \mathcal{A}^-_\mu \mathcal{A}^+_\rho \mathcal{A}^+_\nu \mathcal{A}^-_\lambda - \mathcal{A}^-_\mu \mathcal{A}^+_\lambda \mathcal{A}^+_\nu \mathcal{A}^-_\rho\)
    \] \\
    &+ \mathrm{Perm}(\mu,\nu,\rho,\lambda)\,,
\end{aligned}
\label{eq:L4}
\end{equation}
respectively, with $g^{\mu \nu}$ the space-time metric. Interactions with scalars read as
\begin{equation}
\begin{aligned}
    &g_{_{\mathcal{B} \mathcal{B} h_1 h_1}} = \frac{m_{\mathcal{B}}^2}{v_\varphi^2} \sin^2 \alpha
    \qquad
     g_{_{\mathcal{B} \mathcal{B} h_2 h_2}} = \frac{m_{\mathcal{B}}^2}{v_\varphi^2} \cos^2 \alpha
    \qquad
     g_{_{\mathcal{B} \mathcal{B} h_1 h_2}} = \frac{m_{\mathcal{B}}^2}{v_\varphi^2} \sin(2 \alpha)
     \\
     &g_{_{\mathcal{A} \mathcal{A} h_1 h_1}} = \frac{m_{\mathcal{A}}^2}{v_\varphi^2} \sin^2 \alpha
     \qquad
     g_{_{\mathcal{A} \mathcal{A} h_2 h_2}} =  \frac{m_{\mathcal{A}}^2}{v_\varphi^2} \cos^2 \alpha
     \qquad
     g_{_{\mathcal{A} \mathcal{A} h_1 h_2}} =   \frac{m_{\mathcal{A}}^2}{v_\varphi^2} \sin(2 \alpha)
     \\
     &g_{_{\mathcal{B} \mathcal{B} h_1}} = 2  m_{\mathcal{B}}^2 \sin \alpha
     \qquad
     g_{_{\mathcal{B} \mathcal{B} h_2}} = 2 m_{\mathcal{B}}^2 \cos \alpha
     \qquad
     g_{_{\mathcal{A} \mathcal{A} h_1}} = 2 m_{\mathcal{A}^\pm}^2 \sin \alpha
     \qquad
     g_{_{\mathcal{A} \mathcal{A} h_2}} = 2 m_{\mathcal{A}^\pm}^2 \cos \alpha\,,
\end{aligned}
    \label{eq:self-proca-AB}
\end{equation}
with the scalar mixing angle defined in \cref{eq:rotmat}. { Note that, unlike the pseudo-Goldstone cases discussed above, production of ultralight Proca fields in collision experiments is highly suppressed by their own mass such that, for astrophysically relevant scales, their search at current and next generation of particle colliders is rather challenging if not unrealistic, for the vector DM models presented in this work.}

Most of the features described for the model with a single (real) Proca field are also valid for the current discussion. Among the differences we highlight the possibility of self interactions involving four complex $\mathcal{A}^\pm$. If the size of $m_\mathcal{A}$ results from an extremely small mirror symmetry breaking scale then, the $\U{h.e.m}$ gauge coupling, $e'$, can be sizeable and the self interactions on the last term of \cref{eq:L4} must be considered. On the other hand, if the mass results from a feebly interacting mirror sector, then the theory becomes asymptotically free. A potential observation of Proca stars can be seen as a channel to probe the details of the mirror sector, in particular, to shed light on the scales and interactions strengths involved. The recently suggested identification of GW190521 as a collision of Proca stars~\cite{CalderonBustillo:2020srq} is tentative in this  direction; however the analysis in this work does not consider self-interactions. It would be quite interesting to understand the impact of these. {Notice that, as discussed above, one of the possible ways of making $\mathcal{A}^\pm$ ultralight is by imposing a tiny hidden charge. In such a scenario quartic interactions with the hidden photons $\gamma^\prime$ become extremely suppressed while sizeable self interactions among massive modes (first and last terms in \cref{eq:L4}) can become sizeable in the limit $\theta'_{_W} \to 0$ such that $e^\prime \csc^2 \theta'_{_W} \sim e^\prime \cot^2 \theta'_{_W} \sim \mathcal{O}(1)$. However, not only this is a fine-tuned scenario as well as the masses of both the real and complex Proca fields would become comparable thus affecting the stability of the Proca star. Alternatively, if $e^\prime$ is on its own sizeable, $\mathcal{A}^\pm$ can get annihilated into a pair of $\gamma'$ and the stability of the Proca star needs to take into account strong gravity effects. Such a discussion lies beyond the scope of this manuscript and will be addressed elsewhere.}

Could the real Proca be related to the XENON1T excess?~\cite{XENON:2020rca}~\footnote{The XENON1T experiment has reported results from searches for new physics. An excess over known backgrounds was observed at low energies and most prominent between 2 and 3 keV.} We should be careful because current constraints on bosonic particle mass, in principle, only apply to real fields. However, if the emergence of a complex field implies the existence of real massive fields, those constraints become relevant. Note that, unless the theory is feebly interacting, the first term in \cref{eq:L3} implies that the real Proca field can efficiently decay in a pair of complex ones if $m_{\mathcal{B}} > 2 m_{\mathcal{A}}$. In such a scenario, the complex Proca field can be candidate to, at least, a fraction of the dark matter abundance in the Universe. On the other hand, for $m_{\mathcal{B}} < 2 m_{\mathcal{A}}$, then both Proca fields become dark matter candidates. Note that $h_2$ decay channels to both Proca fields is suppressed either by a tiny scale $v_\varphi \ll m_{h_2}$ or by the hidden electromagnetic gauge coupling $e'$. If the mirror symmetry breaking scale is larger than the EW one with a new visible scalar at colliders, then the mirror sector must be feebly interacting in order to allow for ultralight vector bosons. In such a scenario, only mass terms contribute to the solution of stable Proca stars while self interactions are extremely suppressed. As a final remark, it is worth mentioning that the multiplicity of Proca fields, both complex and scalars, is larger in models invariant under larger hidden gauge symmetries. In fact, the $\SU{2}{H} \times \U{H}$ model introduced here is the minimal scenario where a well motivated complex Proca field in the context of HEP can emerge.

\section{Summary and conclusions}
\label{sec:Conc}


We have presented five simple extensions of the SM capable of providing ultralight real and complex bosons. While fine-tuned solutions are unattractive and potentially problematic from the perspective of Quantum Field Theory, ultralight candidates stemming from sectors constructed upon symmetry arguments are well formulated and stable against quantum corrections. In particular, ultralight scalars can emerge when continuous global symmetries are both explicitly and spontaneously broken whereas Proca fields result from the spontaneous breakdown of new gauge symmetries if either the scale of the theory is well below the EW one or if the interactions strengths are significantly weaker than those of the SM.

We show in \cref{tab:summary} a summary of the models discussed in this article with focus on the nature of the ultralight boson candidates.
\begin{table}
	\centering
	\resizebox{\columnwidth}{!}{%
		\begin{tabular}{cc|cccc|cc} \hline
			Model & Symmetry & Complex Vectors & Real Vectors & Complex Scalars &  Real Scalars & Masses & Self Interactions\\
			\hline
			1 & Global $\U{G}$ & \xmark & \xmark & \xmark  & $\theta,~h_2$ & \eqref{eq:eigvals} & \eqref{eq:self111} \eqref{eq:self112} \eqref{eq:self113} \\
			2 & Global $\U{G} \times \U{G'}$ & \xmark & \xmark & $\eta$  & $h_2,~h_3$  & \eqref{eq:mass123} \eqref{eq:meta} & \eqref{eq:self_eta_U1U1} \eqref{eq:self2} \eqref{eq:self3} \\
			3 & Global $\SU{2}{G} \times \U{G}$ & \xmark & \xmark & $\eta^\pm$  & $\theta_3,~h_2$  & \eqref{eq:mtheta3} \eqref{eq:metapm} \eqref{eq:eigvals2} & \eqref{eq:Vthetas-2} \eqref{eq:self-4-eta} \eqref{eq:self-3-eta} \\
			4 & Local $\U{H}$ & \xmark & $\mathcal{B}$ & \xmark  & $h_2$  & \eqref{eq:mB} & \eqref{eq:self-proca-B} \\
			5 & Local $\SU{2}{H} \times \U{H}$ & $\mathcal{A}^\pm$ & $\mathcal{B}$ & \xmark  & $h_2$  & \eqref{eq:mirror-mass} & \eqref{eq:L3} \eqref{eq:L4} \eqref{eq:self-proca-AB} \\
			\hline
		\end{tabular}%
	}
	\caption{Summary of models. The $h_i$ fields are only viable ultralight candidates in the limit of feebly interacting theories. In model 2 we denote the ultralight complex scalar candidate as $\eta$ instead of $\eta^\pm$ (as in model 3) in order to indicate that the Noether charge is accidental rather than fundamental.
	}
	\label{tab:summary}
\end{table}
In particular, model 3, so far lacking a discussion in the literature, and model 5, introduced here for the first time from the perspective of an ultralight Mirror sector as opposed to a heavier one, can provide extraordinary complex field candidates relevant for boson and Proca stars respectively.

The models presented provide a SM-like Higgs with a mass of 125 GeV with couplings to the remaining SM particles that are either the ones predicted by the SM or are such that 
the SM limit can be attained in a simple way. Together with the SM Higgs, there are in many cases more scalars that can be lighter or heavier than the SM Higgs but are expected to have their mass at the electroweak scale. In fact, many of these extensions, although not the ultralight scalars, have been probed in the previous LHC runs and searches will continue at next run. A wealth of phenomenological information combining astrophysical sources, collider data and cosmological observations can potentially shed light on the existence of a new ultralight bosonic sector.

\bigskip

\acknowledgments

This work is supported by the Center for Research and Development in Mathematics and Applications (CIDMA) through the Portuguese Foundation for Science and Technology (FCT - Funda\c c\~ao para a Ci\^encia e a Tecnologia), references UIDB/04106/2020 and, UIDP/04106/2020, and by national funds (OE), through FCT, I.P., in the scope of the framework contract foreseen in the numbers 4, 5 and 6 of the article 23, of the Decree-Law 57/2016, of August 29, changed by Law 57/2017, of July 19. This work is also supported by CFTC-UL through FCT, references UIDB/00618/2020 and UIDP/00618/2020.
We acknowledge support  from  the  projects PTDC/FIS-OUT/28407/2017, PTDC/FIS-PAR/31000/2017, CERN/FIS-PAR/0027/2019, CERN/FIS-PAR/0002/2019 and PTDC/FIS-AST/3041/2020. This work has further been supported by the European Union’s Horizon 2020 research and innovation (RISE) programme H2020-MSCA-RISE-2017 Grant No. FunFiCO-777740.  
This work has also been supported in part by the Swedish Research Council grant, contract number 2016-05996 and by the European Research Council (ERC) under the European Union's Horizon 2020 research and innovation programme (grant agreement No 668679).
The authors would like to acknowledge networking support by the COST Action CA16104 and from the HARMONIA project, contract UMO-2015/18/M/ST2/0518.

\appendix

\section{Quartic and cubic self interactions for the global $\U{G} \times \U{G'}$ model}
\label{app:self1}

We give in this Appendix a list of quartic and cubic self interactions involving all ultralight candidate bosons, $\eta$. $h_2$ and $h_3$. Recall that, unlike $\eta$, $h_2$ and $h_3$ can be also be heavier than the EW scale. 
\begin{equation}
    \begin{aligned}
        & \lambda_{_{h_2 h_2 h_2 h_2}} = \frac{1}{16} \[ (\lambda_{12} + \lambda_\phi) \cos^4 \alpha_h + 2 \lambda_H \sin^4 \alpha_h + \lambda_{H \phi} \sin^2 (2 \alpha_h) \]
        \\
        & \lambda_{_{h_3 h_3 h_3 h_3}} = \frac{1}{16} (\lambda_{12} + \lambda_\phi)
        \\
        & \lambda_{_{h_2 h_2 h_3 h_3}} = \frac{1}{8} \[ (3 \lambda_\phi - \lambda_{12}) \cos^2 \alpha_h + 2 \lambda_{H \phi} \sin^2 \alpha_h \]
        \\
        & \lambda_{_{h_1 h_1 h_2 h_2}} = \frac{1}{64} \[ 3 \lambda_{12} + 6\lambda_H + 4 \lambda_{H\phi} + 3\lambda_\phi - 3 \( \lambda_{12} + 2 \lambda_H - 4 \lambda_{H \phi} + \lambda_\phi \) \cos (4 \alpha_h)\]
        \\
        & \lambda_{_{h_1 h_1 h_3 h_3}} = \frac{1}{8} \[ 2 \lambda_{H\phi} \cos^2 \alpha_h - (\lambda_{12} - 3 \lambda_\phi) \sin^2 \alpha_h \]
        \\
        & \lambda_{_{h_1 h_1 \eta \bar{\eta}}} = \frac{1}{4} \[ 2 \lambda_{H\phi} \cos^2 \alpha_h + (\lambda_{12} + \lambda_\phi) \sin^2 \alpha_h \]
        \\
        & \lambda_{_{h_2 h_2 \eta \bar{\eta}}} = \frac{1}{4} \[ 2 \lambda_{H\phi} \sin^2 \alpha_h + (\lambda_{12} + \lambda_\phi) \cos^2 \alpha_h \]
        \\
        & \lambda_{_{h_3 h_3 \eta \bar{\eta}}} = \frac{1}{4} (\lambda_{12} + \lambda_\phi)
        \\
        & \lambda_{_{h_1 h_2 \eta \bar{\eta}}} = \frac{1}{4} \( 2 \lambda_{H\phi} - \lambda_{12} - \lambda_\phi  \) \sin (2 \alpha_h)\,,
    \end{aligned}
    \label{eq:self2}
\end{equation}
and
\begin{equation}
    \begin{aligned}
    & \lambda_{_{h_1 \eta \bar{\eta}}} = v_h \lambda_{H\phi} \cos \alpha_h + \frac{1}{\sqrt{2}} v_\sigma (\lambda_{12} +\lambda_\phi ) \sin \alpha_h
    \\
    & \lambda_{_{h_2 \eta \bar{\eta}}} =  v_h \lambda_{H\phi} \sin \alpha_h - \frac{1}{\sqrt{2}} v_\sigma (\lambda_{12} +\lambda_\phi ) \cos \alpha_h
    \\
    & \lambda_{_{h_3 \bar{\eta} \bar{\eta}}} = \lambda_{_{h_3 \eta \eta}} =  \frac{1}{2\sqrt{2}} v_\sigma (\lambda_{12} -\lambda_\phi ) \cos \alpha_h
    \\
    & \lambda_{_{h_1 h_2 h_2}} = \frac14 \left\{ 2 v_h \lambda_{H\phi} \cos^3 \alpha_h + \sqrt{2} v_\sigma (3 \lambda_{12} - 4 \lambda_{H \phi} + 3\lambda_\phi ) \cos^2 \alpha_h \sin \alpha_h \right.
    \\
    &
    \left.
    \phantom{\lambda_{_{h_1 h_2 h_2}} = \frac14 }
    + 2 \sin^2 \alpha_h \[v_h (3 \lambda_H - 2 \lambda_{H \phi}) \cos \alpha_h 
    +\sqrt{2} v_\sigma \lambda_{H \phi} \sin \alpha_h
    \] \right\}
    \\
    & \lambda_{_{h_1 h_3 h_3}} = \frac14 \[ 2 v_h \lambda_{H\phi} \cos \alpha_h - \sqrt{2} v_\sigma (\lambda_{12} - 3\lambda_\phi ) \sin \alpha_h \]
    \\
    & \lambda_{_{h_2 h_3 h_3}} = \frac14 \[ 2  v_h \lambda_{H\phi} \sin \alpha_h + \sqrt{2} v_\sigma (\lambda_{12} - 3\lambda_\phi ) \cos \alpha_h \]
    \\
    & \lambda_{_{h_2 h_2 h_2}} = \frac14 \[ 2 v_h \sin \alpha_h \( \lambda_{H \phi} \cos^2 \alpha_h + \lambda_H \sin^2 \alpha_h\) \right.
    \\
    &
    \left.
    \phantom{\lambda_{_{h_2 h_2 h_2}} = \frac14 }
    - \sqrt{2} v_\sigma \cos \alpha_h \( 2 \lambda_{H \phi} \sin^2 \alpha_h + (\lambda_{12} + \lambda_\phi) \cos^2 \alpha_h \) \]\,.
    \end{aligned}
    \label{eq:self3}
\end{equation}
The quartic couplings are expressed in terms of the physical masses, the mixing angle $\alpha_h$ and the VEVs as
\begin{equation}
    \begin{aligned}
        &\lambda_H = \frac{m_{h_1}^2 \cos^2 \alpha_h + m_{h_2}^2 \sin^2 \alpha_h }{v_h^2} \qquad ~~~~~~~~
        \lambda_{H \phi} = \frac{\( m_{h_1}^2 - m_{h_2}^2 \) \cos \alpha_h \sin \alpha_h}{\sqrt{2} v_h v_\sigma}
        \\
        &\lambda_\phi = \frac{m_{h_1}^2 \sin^2 \alpha_h + m_{h_2}^2 \cos^2 \alpha_h + m_{h_3}^2}{2 v_\sigma^2}
        \qquad
        \lambda_{12} = \frac{m_{h_1}^2 \sin^2 \alpha_h + m_{h_2}^2 \cos^2 \alpha_h - m_{h_3}^2}{2 v_\sigma^2}\,.
    \end{aligned}
\end{equation}

\bibliographystyle{JHEP}
\bibliography{biblio}

\end{document}